\documentclass[twocolumn]{autart}    
\usepackage{graphicx,color}
\usepackage{amsfonts}   
\usepackage{amsmath}    
\usepackage[norelsize,algo2e,ruled,linesnumbered]{algorithm2e}    
\usepackage{amssymb}
\usepackage{float}     
\usepackage{mathtools} 
\usepackage{subcaption} 
\usepackage[title]{appendix}
\usepackage{array}
\usepackage{diagbox}
\usepackage[justification=centering]{caption}

\usepackage{mathtools}
\mathtoolsset{showonlyrefs=true}

\makeatletter
\renewcommand*{\@opargbegintheorem}[3]{\trivlist
  \item[\hskip \labelsep{\bfseries #1\ #2}] \textbf{(#3)}\ \itshape}
\makeatother

\newcommand{\red}[1]{\textcolor{red}{#1}}

\begin{document}

\begin{frontmatter}
\runtitle{Differential Privacy for Symbolic Systems}  
\title{Differential Privacy for Symbolic Systems \\ with Application to Markov Chains\thanksref{footnoteinfo}} 
\thanks[footnoteinfo]{This paper was not presented at any IFAC meeting. Corresponding author: Bo Chen. BC and MH were supported by NSF CAREER Grant No. 1943275, ONR Grant N00014-21-1-2502, and
AFOSR Grant FA9550-19-1-0169.}
\thanks{Kevin Leahy is currently an MIT Lincoln Laboratory employee. No Laboratory funding or resources were used to produce the result/findings reported in this publication.
}

\author[UF]{Bo Chen}\ead{bo.chen@ufl.edu},    
\author[MIT]{Kevin Leahy}\ead{kevin.leahy@ll.mit.edu}, 
\author[LL]{Austin Jones}\ead{austinmjonesresearch@gmail.com},               
\author[UF]{Matthew Hale}\ead{matthewhale@ufl.edu} 

\address[UF]{Department of Mechanical and Aerospace Engineering, University of Florida, Gainesville, FL, USA.}  
\address[MIT]{Massachusetts Institute of Technology Lincoln Laboratory, Lexington, MA, USA.}             
\address[LL]{No affiliation. Contributions made while this author was affiliated with MIT Lincoln Laboratory, Lexington, MA, USA.}
          
                                           
\begin{abstract}                          
Data-driven systems are gathering increasing amounts of data from users, and sensitive user data requires privacy protections. 
In some cases, the data gathered is non-numerical or symbolic, and conventional approaches to privacy, e.g., adding noise, do not apply, though
such systems still require privacy protections. 
Accordingly, we present a novel differential privacy framework for protecting trajectories generated by symbolic systems. 
These trajectories can be represented as words or strings over a finite alphabet. 
We develop new 
differential privacy mechanisms that approximate a sensitive word using a random word that is likely to be near it. 
An offline mechanism is implemented efficiently using a Modified Hamming Distance Automaton to generate 
whole privatized output words over a finite time horizon. Then, an online mechanism is implemented by taking in  
a sensitive symbol and generating a randomized output symbol at each timestep. 
This work is extended to Markov chains to generate differentially private state sequences that a given
Markov chain could have produced. 
Statistical accuracy bounds are developed to quantify the accuracy of these mechanisms, and
numerical results validate the accuracy of these techniques for 
strings of English words. 
\end{abstract}
\end{frontmatter}

\section{Introduction}
As control applications have become increasingly reliant on user data, there has arisen interest in 
protecting individuals’ privacy, e.g., in smart power grids~\cite{SiddiquiGrid2012,SimmhanGrid2011} and smart transportation systems~\cite{bloom2017self,Hassan2020}. 
Researchers have proposed various quantitative definitions of privacy, and the notion of differential 
privacy has emerged as one standard privacy specification in recent years~\cite{CortesPrivacySummary2016,dwork08}. 
The statistical nature of differential privacy makes it unlikely for an eavesdropper or adversary to 
learn anything meaningful about sensitive data from its differentially private form. Its key features 
include immunity to post-processing~\cite{Dwork2014}, in that transformations of privatized data do not 
weaken privacy guarantees, and robustness to side information, in that learning additional information 
about data-producing entities does not substantially
weaken differential privacy~\cite{Kasiviswanathan_Smith_2014}.
Immunity to post-processing means that differentially private data can be freely used without harming
privacy's guarantees. And there is a growing body of work on differential privacy in systems and control that exploits this property, 
including
in multi-agent control~\cite{CortesPrivacySummary2016,Hawkins2020,wang17}, convex optimization~\cite{Han2017,Hale2015,Nozari2016,Huang2015}, 
linear-quadratic control~\cite{hale2019privacy,Koufogiannis2017}, 
controller design~\cite{kawano20}, and 
filtering and estimation problems~\cite{Ny2014,Ny2018}.
These works implement differential privacy 
for numerical data using the Laplace or Gaussian mechanisms, 
which add noise to sensitive data before sharing it.

Symbolic control systems generate sequences of non-numerical data, 
            which can often be represented as words or strings over a finite 
            alphabet. 
            A symbolic trajectory can represent, for example, a sequence of modes to switch between 
            in a hybrid system~\cite{Stiver1992,BARTON20061576}, a sequence of regions of state space to occupy, e.g., in a path planning problem~\cite{belta07},
            or the statuses of smart devices in the Internet of Things~\cite{Spiess2009,Li2018}. 
            In these and other applications, symbolic systems require privacy protections
            just as their numerical counterparts do. For example, the locations of a patrolling robot
            over time may require protections in order to prevent an adversary from predicting
            where the robot will go next. 
            In addition, symbolic trajectories can be produced by
            Markov chains and Markov decision processes, which have been
            used to model~traffic systems~\cite{rong16}, smart buildings~\cite{Akkaya2015}, cyber physical systems~\cite{Li2015}, and robot navigation systems~\cite{Koenig1998}.
            Symbolic trajectories produced by 
            these systems can reveal the locations of a user, their activities, and their acquaintances,
            yet it is desirable to share such trajectories to enable multi-agent coordination,
            learning from trajectory data, and other data-driven applications. Therefore, we
            develop novel differential privacy protections that enable privatized symbolic data to be shared,
            and our developments apply to these applications and any others in which symbolic
            data contains sensitive information, including any applications modeled as Markov chains. 
            

Non-numerical data typically cannot be privatized with additive noise, but 
it can be kept differentially private using the exponential mechanism~\cite{Dwork2014}. 
 The exponential mechanism takes non-numerical data as input and randomly outputs non-numerical data
based on its ``quality'', which is user-specified. 
There is not an inherent notion of quality for symbolic systems,
and in this work we choose to use the Hamming distance as the notion of quality,
which means that a private word is of high quality if it is close to the
sensitive word it approximates. 
It is known that the 
 exponential mechanism can have high computational 
 complexity when its input and output spaces
 are large~\cite{Dwork2014}.
 Indeed, 
 Section~\ref{sec:background_exp} shows that the computational complexity of 
 a na{\"i}ve implementation for strings would be exponential
 in the length of those strings. 
 Thus, more tractable approach is required. 
 
Accordingly, this paper develops an efficient method for the privatization of sensitive words generated by symbolic systems. 
The first contribution of this paper is the definition of differential privacy itself in this context.
Given a sensitive input word, differential privacy requires the generation of private output words
that are near the input word (in the Hamming sense) with high probability. 
The second contribution is a computationally efficient privacy mechanism that constructs
and uses a
modified Hamming Distance nondeterministic finite state automaton (NFA) to generate private output words. 
This mechanism operates on whole strings offline.

In some applications, real-time reporting of status information is necessary for effective operation, e.g., 
real-time IoT monitoring systems~\cite{Zhou2019} and smart home management systems~\cite{Cui2014}. 
Also, in cloud control~\cite{Lofaro2015},
states are transmitted to an aggregator and the aggregator sends back commands in real-time.
To provide privacy in such applications, the third contribution of this work 
is an online differential privacy
mechanism. It generates individual random symbols in a way that privatizes entire symbolic 
trajectories. 
This mechanism has the advantage that future symbolic states are not needed \emph{a priori}.

The fourth contribution is the extension of these differential privacy mechanisms to Markov chains. 
Markov processes specifically have been considered in 
applications such as traffic systems~\cite{Yin2013}
and healthcare~\cite{Zois2016}. 
These system setups require the sharing of symbolic trajectories that can be sensitive 
because they may contain a user's destination or personal health information. 
Each 
sensitive symbol in a Markov chain trajectory 
is dependent on its previous symbol, and only some symbol-to-symbol transitions are feasible. 
Thus, both
the offline and online mechanisms are modified 
to generate output trajectories that
are feasible with respect to the dynamics of a given Markov chain. 

Concentration bounds are developed for the accuracy of each mechanism, which 
enable the calibration of privacy 
protections based on the acceptable error in a given application. 
These mechanisms 
are demonstrated on a Markov chain generated by the traffic data for some of the major streets in Gainesville, Florida, which is available at Florida Traffic Online (2021)~\cite{florida2021}.


A preliminary version of this work appeared in~\cite{jones2018differential}. The current paper extends this work in three ways. First, we provide a full proof that the mechanism in~\cite{jones2018differential} provides
differential privacy, and we also derive novel error bounds for it. 
Second, we provide a new differential privacy mechanism for the online setting,            
which can be implemented for real-time control, and accuracy bounds are provided for it as well. Third, both the offline and online mechanisms are extended to Markov chains and 
two new privacy mechanisms are developed. Both 
are shown
to provide differential privacy while ensuring that all privately generated words
are feasible with respect to the dynamics of the underlying Markov chain. Error bounds
are also presented for this mechanism.


 
 Other approaches to protecting information include \emph{opacity}~\cite{Saboori2014}, which
bounds the probability of correct state estimates
by an adversary that observes all actions
in an MDP~\cite{Wu2018,Wu2018Insertion}.
Our paper differs by considering problems in which
a user chooses to share information privately, rather
than maintaining secrecy under observation.

A related
body of work introduced \emph{approximate opacity}~\cite{Yin2021}, which 
studies the protection of the states of a system while an intruder makes approximate
observations of that system. This work differs by considering a setting in which
differentially private state observations are deliberately shared with an outside
party, rather than being the product of observations of a system. That is, 
while approximate opacity studies inherent system properties that make it difficult
to infer the system's states, we develop new private output maps that
can be added to an existing system to enforce differential privacy. Additionally, in~\cite{savas2018entropy,Savas2020}, policies are synthesized to minimize the predictability of trajectories to an outside observer 
 by maximizing the entropy of MDPs. Conversely,
 we develop private output maps that do
 not require re-synthesizing a policy for privacy. 
  In addition, our approaches enforce differential privacy at the trajectory level,
 rather than pointwise in time~\cite{ramasubramanian2020privacypreserving}.

The rest of the paper is organized as follows. Section~\ref{sec:preliminaries} presents background, and 
Section~\ref{sec:prob_statement} gives formal problem statements. 
Section~\ref{sec:dp_nonmarkov_chain} develops privacy mechanisms, and 
Section~\ref{sec:dp_markov_chain} applies them to Markov chains. 
Section \ref{sec:experiment} gives numerical results, and Section~\ref{sec:conclusion} concludes.

\noindent \textbf{Notation} 
Let $\mathbb{N}$ denote the set of all non-negative integers and $\mathbb{N}^+$ denote the the set of all positive integers. For~$n \in \mathbb{N}^+$, let~$[n]=\{1,\dots,n\}$. 
$\Sigma$ denotes a finite alphabet. 
A word of length~$n$ over~$\Sigma$ is a concatenation of 
symbols~$w = \sigma_1\sigma_2 \dots \sigma_n$ with~$\sigma_i\in\Sigma$ for 
all~$i \in [n]$. 
We also write~$w_i$ for the~$i^{th}$ symbol
in the word~$w$. 
Let~$\Sigma^n$
denote all words of length~$n$ over~$\Sigma$, and $2^\Sigma$ denote the power set of $\Sigma$.
The notation~$|w|$ denotes the length of a word~$w$. 

\section{Preliminaries on Symbolic Systems}\label{sec:preliminaries}
A finite state automaton (FSA) 
is a tuple $A=(Q,\Sigma,q^0,\delta,F)$, where $Q$ is a set of states, $\Sigma$ is an input 
alphabet, $q^0\in Q$ is the initial state, $\delta: Q\times\Sigma \to Q$  is the transition 
function between states, and $F\subseteq Q$ is the set of accepting states. If the transition 
function~$\delta$ is a nondeterministic mapping, i.e. $\delta: Q\times\Sigma \to 2^Q$, then this FSA is 
called a nondeterministic finite state automaton (NFA).

Given an NFA~$A=(Q,\Sigma,q^0,\delta,F)$, a word~$w=\sigma_1\sigma_2\dots\sigma_n$, with~$\sigma_i \in \Sigma$, induces
a \emph{run}, which is 
a word~$q=q_0q_1\dots q_n\in Q^*$ such that~$q_0=q^0$ and~$q_{i+1}\in\delta(q_i,\sigma_{i+1})$. 
The automaton~$A$ \emph{accepts} a word~$w_o$ if the final state of the induced run is an accepting state, i.e.,~$q_n\in F$. 
The set of words accepted by~$A$ is its \emph{language}, denoted~$\mathcal{L}(A)$.

To compare two words, we introduce the Hamming distance. 
\begin{defn}[Hamming Distance]
Given an alphabet $\Sigma$, for two $n$-length words $v ,w \in\Sigma^n$, the Hamming distance between them, denoted $d(v ,w)$, is the number of positions at which the corresponding symbols are different. Mathematically, we have~$d(v, w) = \big|\{i \mid v_{i} \neq w_{i}\}\big|$, where~$\big|\cdot\big|$
denotes cardinality. 
\end{defn}
In other words, the Hamming distance is a metric that measures the minimum number of substitutions that can be applied to $v$ to convert it to $w$.
For example, the Hamming distance between ``hammer'' and ``bumper'' is~$3$.

\section{Privacy Background and Problem Statements} \label{sec:prob_statement}
This section gives background and the problem statements that are the focus of the remainder of the paper. 

\subsection{Basic Differential Privacy Definitions}\label{sec:diff_privacy_description}
Differential privacy is enforced by a \emph{mechanism}, which is a randomized map. 
For similar pieces of sensitive data, a mechanism must produce outputs that are 
approxmiately distinguishable. The definition of ``similar''  
is given by an adjacency relation, 
which takes the following form
for words generated by a symbolic system. 

\begin{defn}[Word Adjacency]\label{def:word_adjacency}
    Fix a length~$n \in \mathbb{N}^+$ and an adjacency parameter~$k \in \mathbb{N}$. 
    The word adjacency relation on~$\Sigma^n$
    is
        $\textnormal{Adj}_{n,k}=\big\{(w_1,w_2) \mid d(w_1,w_2)\leq k\big\}$. 
\end{defn}

\begin{figure}
\centering
\includegraphics[width=.30\textwidth]{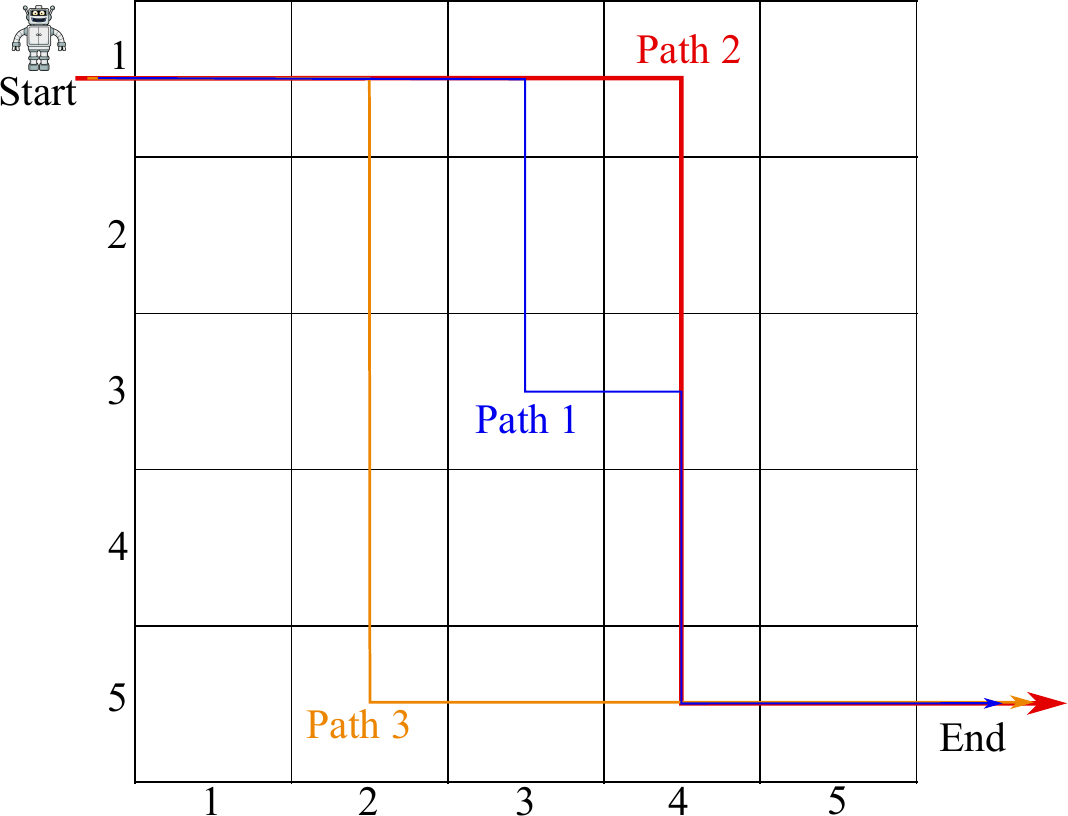}
\caption{A robot is traveling through a $5\times 5$ grid. It may choose from three paths, and each path can be represented as a word consisting of the coordinates it passes through.
}
\label{fig:example_adjacency}
\end{figure}

Two words in~$\Sigma^n$ are adjacent if the Hamming distance between them is no more than~$k$. 
For example, consider Figure~\ref{fig:example_adjacency}, where the robot can travel from~$(1,1)$ to~$(5,5)$ 
along Path~$1$,~$2$, or~$3$. Each path gives a length~$9$ word of grid cells the 
robot traverses. With~$k=3$, Paths~$1$ and~$2$ are 
word adjacent (since they differ in two points), 
but Path~$3$ is not word adjacent to the other two (due to differing in more than 3 points).  
Adjacent words must be made approximately indistinguishable, and~$k$
is the size of difference that must be masked. 

We define adjacency in terms of the Hamming distance because this allows
the mechanisms we develop to protect sensitive differences between trajectories.
For example, suppose a vehicle unexpectedly deviates from its nominal commute
and that we would like to mask this deviation using differential privacy. 
If the nominal trajectory and the
deviating trajectory are adjacent to each other, then differential privacy 
will render the deviating trajectory approximately
indistinguishable from the nominal trajectory, thereby concealing
the sensitive deviation. 
It is precisely these protections that we attain 
by defining adjacency in terms of the Hamming distance between
such trajectories. We note as well that differential privacy for
state space systems (with numerical trajectories) defines adjacency
in terms of an appropriate metric in an~$\ell_p$-space~\cite{Ny2014}, and
thus our use of the Hamming distance is the natural analog of
existing work for the symbolic setting.

We next introduce the definition of differential privacy for symbolic systems we use throughout this paper.

\begin{defn}[Word Differential Privacy]
    Fix a probability space $(\Omega, \mathcal{F}, \mathbb{P})$, an adjacency parameter~$k \in \mathbb{N}$, 
    a length~$n \in \mathbb{N}^+$, and a privacy parameter~$\epsilon > 0$. 
    A mechanism $M_w:\Sigma^n\times\Omega\to\Sigma^n$ is word $\epsilon$-differentially private if, 
    for all words $(w_1,w_2)\in\textnormal{Adj}_{n,k}$ and all~$L \subseteq \Sigma^n$, 
    it satisfies
        $\mathbb{P}\big[M_w(w_1)\in L\big] \leq e^\epsilon\mathbb{P}\big[M_w(w_2)\in L\big]$.
\end{defn}

The parameter~$\epsilon$ sets the strength of privacy protections, and
smaller~$\epsilon$ implies stronger privacy. 
In the literature, $\epsilon$ typically ranges from $0.01$ to~$10$~\cite{Hsu_2014}.
A word differential privacy mechanism guarantees that the randomized outputs of two~$k$-adjacent 
words will be made
approximately indistinguishable to any recipient of their privatized forms, 
including any eavesdroppers. Thus, these recipients are unlikely to 
determine the underlying sensitive word or make high-confidence inferences about it. 


We note that other metrics can be defined on the set of words that allow
for the comparison of words of different lengths. One example is the Levenshtein
distance~\cite{levenshtein65}, 
which allows substitutions of symbols between words, as well as insertions
and deletions. However, we do not use such a metric for 
adjacency because doing so would allow a privacy mechanism to
generate private outputs whose lengths are different from the lengths
of the inputs that produce them. This can be problematic, particularly
when a private output is shorter than the sensitive input that produced it.
To see why, consider a symbolic trajectory that captures the sequence
of intersections traversed by an autonomous vehicle. If a privacy mechanism
deletes entries of such a trajectory, then it has deleted data about the
location history of that vehicle. In doing so, the privacy mechanism can
generate semantically invalid data, e.g., a sequence of intersections
that cannot actually be consecutively followed from the starting point to the ending point. 
Such instances motivate us to use the Hamming distance in this work because it
can only allow words to be adjacent if they have the same length.


\subsection{Differential Privacy Problem Statements}
Formal privacy problem statements are given next. 

\begin{prob}\label{prob:offline_mechanism}
    Fix a probability space $(\Omega, \mathcal{F}, \mathbb{P})$. Given a word length~$n \in \mathbb{N}^+$, an alphabet $\Sigma$, 
    an adjacency parameter~${k \in \mathbb{N}}$, and an adjacency relation $Adj_{n,k}$, develop an offline
    word $\epsilon$-differential privacy mechanism $M_w^{\text{off}}:\Sigma^n \times \Omega \to \Sigma^n$ 
    that takes as input an entire word~$w$ that is already generated, i.e., for all $i\in[n]$, $\sigma_i$ is known a priori.
\end{prob}

Problem~\ref{prob:offline_mechanism} mathematically formulates an \emph{offline} mechanism, in the sense
that it privatizes an entire word after it has been generated. This would be used when data is first harvested,
then privatized and released in batches. 

We are also interested in the \emph{online} setting to account 
for cases in which a symbolic trajectory must be privatized and shared as it is 
generated. 
One symbol is shared at each point in time, and thus~$n$ is both the length of a word
and the length of time horizon over which it is shared. 
For a word $w=\sigma_1\sigma_2\dots\sigma_n$, the online setting 
shares~$\sigma_t$ for each time $t\in[n]$, which leads to the following. 

\begin{prob}\label{prob:factored_mechanism}
    Fix a probability space $(\Omega, \mathcal{F}, \mathbb{P})$. Given a word length~$n \in \mathbb{N}^+$, an alphabet $\Sigma$,
    an adjacency parameter~${k \in \mathbb{N}}$, and an adjacency relation $Adj_{n,k}$, develop
    an online mechanism $M_w^{\text{on}}:\Sigma^n\times\Omega\to\Sigma^n$
    that is word~$\epsilon$-differentially private,
    where only the symbols~$\sigma_{1}\cdots\sigma_{t}$ of $w$ are known at each time $t$, i.e., the future symbols $\sigma_{t+1}\cdots\sigma_n$ after time~$t$ are unknown.
\end{prob}

For a sensitive word~$w = \sigma_1\sigma_2\dots\sigma_n\in\Sigma^n$, the online mechanism $M_w^{\text{on}}$ 
approximates the sensitive symbol~$\sigma_t$ at each $t\in[n]$ with a random symbol. 
The challenge is that, for any time $t\in[n]$, $M_w^{\text{on}}$ only has access to symbols generated 
before time $t$, while differential privacy must be enforced for the entire word. 
Thus, the symbol-by-symbol randomization of~$M_w^{\text{on}}$ must enforce the correct
distribution over entire words without knowing the entire words. 
Problems~\ref{prob:offline_mechanism} and~\ref{prob:factored_mechanism} are solved in Section~\ref{sec:dp_nonmarkov_chain}.

\subsection{Extension to Markov Chains}
A widely used class of symbolic system models is Markov chains. A Markov chain is a sequence of 
random variables $X_1,X_2,\dots\in S$ with the Markov property, i.e., 
the value of~$X_{t+1}$ depends only on the value of~$X_t$. 
The state space~$S$ contains all possible values
of $X_t$. The transition probability of going from state $s_i$ 
to state~$s_j$ in one step is~$p_{i,j}=\mathbb{P}[s_j \mid s_i]$. 
The matrix of transition probabilities is denoted~$P$,
where~$P_{ij} = p_{i,j}$. In this work, the tuple~$(S, P, s_0)$
denotes the Markov chain with state space~$S$, transition matrix~$P$,
and initial state~$s_0 \in S$. 
A state $s_j$ is called a \emph{feasible state} 
of another state $s_i$ if $\mathbb{P}[s_j \mid s_i]>0$. 
For a Markov chain with state space $S$, let $S^*$ denote all sequences it
can generate in finite time and $S^n$ denote all sequences that is in $n$ length. Any such sequence over a finite horizon~$[n]$ can be identified with
a word $w=s_0s_1\dots s_n\in S^n$. If for all $t\in[n-1]$, $\mathbb{P}[s_{t+1} \mid s_t]>0$, i.e., 
every state is a feasible state of its previous state, then the word $w$ is called \emph{feasible}.
The set of feasible words of length $n$ is denoted~$\mathcal{L}(S^n)$.

For privacy of state sequences of 
Markov chains, the goal is to generate a private sequence of states which is 
feasible with respect to its dynamics. 


\begin{prob}\label{prob:markov_chain}
    Fix a probability space $(\Omega, \mathcal{F}, \mathbb{P})$. Given a word length~$n \in \mathbb{N}^+$, an alphabet $\Sigma$, 
    an adjacency parameter~${k \in \mathbb{N}}$, and an adjacency relation $Adj_{n,k}$, for a Markov chain~$(S, P, s_0)$,
    develop word $\epsilon$-differentially private offline and online mechanisms,~$M_{w,s}^{\text{off}}$ and~$M_{w,s}^{\text{on}}$, 
    such that each output word is feasible with respect to the allowable transitions of the Markov chain.
    The offline mechanism takes in an entire word~$w$, while the online mechanism only has access to 
    the symbols~$\sigma_{1}\cdots\sigma_{t}$ at each time $t$, i.e., the future symbols $\sigma_{t+1}\cdots\sigma_n$ after time~$t$ are unknown.
\end{prob}

We note here that Problems~\ref{prob:offline_mechanism},~\ref{prob:factored_mechanism},
and~\ref{prob:markov_chain} merely state the privacy requirements of each setting,
but they deliberately do not constrain the allowable implementations of the privacy
mechanisms that solve them. For example, in Problem~\ref{prob:factored_mechanism},
a privacy mechanism may have some form of internal
memory so that each output symbol depends on the input string and past output
symbols. This setup is permitted despite not being a requirement in
Problem~\ref{prob:factored_mechanism}. Indeed, a solution to each problem requires
only some mechanism that enforces differential privacy and maps between the
appropriate spaces, regardless of its internal implementation details.

Our study of Markov chains is motivated by
our interest in the development of trajectory-level
privacy for symbolic systems, analogous to the
notion of trajectory-level privacy
developed for state space systems in~\cite{Ny2014}. 
In that work, stochastic systems are studied,
which means that privacy at the trajectory
level does not collapse to privacy of 
the initial state of the system (as would
be the case for deterministic systems, e.g.,~\cite{han18,Justin2016,Han2014,Huang2012,hsu2014privately}).
Similarly, in this work we do not consider
deterministic finite-state automata because
their determinism does cause such a collapse
of privacy. Instead, we consider Markov chains
because their stochastic dynamics make
it meaningful to define privacy at the
level of symbolic trajectories. 

\subsection{Background: Exponential Mechanism} \label{sec:background_exp}
This section briefly reviews the exponential mechanism, 
which is the foundation for the privacy mechanisms we develop
(Section 3.4 of~\cite{Dwork2014} gives background).
We emphasize that this paper does not merely implement the exponential mechanism.
Instead, as described in Remark~\ref{rem:complexity} below, the exponential mechanism would be computationally prohibitive to use in its standard form, 
and this work must develop significantly less computationally
complex mechanisms.
This subsection simply gives the privacy goals
that motivate this work, and actually attaining those goals will be the focus of this work. 

For a non-numerical query~$f$ with range~$\mathcal{R}$, the exponential mechanism is implemented by first 
computing the query $f$ on a given input $w_i$, and then sampling a random output 
from~$\mathcal{R}$ that suitably approximates~$f(w_i)$. 
Mathematically, 
the probability of selecting an output depends on its utility score, 
and outputs with high utility scores are generated with high probability. 
In this work, the query under consideration is the identity query, i.e.,~$f(w_i) = w_i$, which
means that we must randomize an entire symbolic word, rather than some function of it. 
The definition of utility score is a design choice made in developing an exponential mechanism, and, given
the use of the identity query, this work uses the following.

\begin{defn}[Utility function]\label{def:utility_function}
    Fix an alphabet~$\Sigma$ and a length~$n \in \mathbb{N}^+$, and consider a sensitive input word~$w_i \in \Sigma^n$. 
    Then a private output word~$w_o \in \Sigma^n$ provides utility equal 
    to~$u(w_i,w_o) = -d(w_i,w_o)$. 
\end{defn}

This utility function encodes the fact that~$w_o$ is a better output for~$w_i$ when it is close to~$w_i$. 
The probability distribution used by the exponential mechanism depends on both the value of~$u$ and its sensitivity, defined next. 

\begin{lem}[Sensitivity~\cite{jones2018differential}] \label{lemma:sensitivity}
    Fix an alphabet~$\Sigma$, a length~$n \in \mathbb{N}^+$, a set~$L \subseteq \Sigma^n$, 
    and 
    an adjacency parameter~$k \in \mathbb{N}$. Then the sensitivity of~$u$ is
    \begin{equation}
        \Delta u = \max_{v\in L}\max_{\substack{w_1,w_2\in L\\(w_1,w_2)\in\textnormal{Adj}_{n,k}}} |u(w_1,v)-u(w_2,v)| \leq k.
    \end{equation}
\end{lem}
The sensitivity bounds the amount by which~$u$ can differ between two 
adjacent input words in~$L$. One can have~$L = \Sigma^n$, though Lemma~\ref{lemma:sensitivity}
also allows for~${L \subsetneq \Sigma^n}$ 
for systems in which some words are infeasible. 

\begin{defn}[Exponential Mechanism]\label{def:exponential_mechanism}
    Fix an alphabet $\Sigma$, a length~$n \in \mathbb{N}^+$, a language $L\subseteq\Sigma^n$, and an adjacency parameter~$k \in \mathbb{N}$. 
    For a sensitive input word $w_i\in L$, the exponential mechanism $M_e$ 
    outputs~$w_o \in L$ with probability 
    \begin{align}
        p_e(w_o) \!&=\! \frac{\exp\left(\frac{\epsilon u(w_i,w_o)}{2\Delta u}\right)}{\sum\limits_{w'\in L} \!\!\exp\left(\frac{\epsilon u(w_i,w')}{2\Delta u}\right)} \!=\! K_w\exp\left(\!\frac{\epsilon u(w_i,w_o)}{2\Delta u}\!\right). 
    \end{align}    
\end{defn}
\begin{rem}
    For two private output words~$w_1$ and $w_2$ and an input word $w_i$, if $d(w_i,w_1)=d(w_i,w_2)$, then then exponential mechanism will output them with equal probability. This is an essential property that we will use to design 
    efficient privacy mechanisms in later sections.
\end{rem}
\begin{rem} \label{rem:complexity}
As stated, this would provide word~$\epsilon$-differential privacy, though
directly implementing the exponential mechanism over $\Sigma^n$ is infeasible.
For a fixed $w_i$, to determine the proportionality constant $K_w$, one would need to compute the Hamming distance from~$w_i$ to 
every possible output word. 
There are~$m^n$ total strings of length~$n$ 
on an alphabet of~$m$ symbols, and 
the time complexity of computing the required distances is~$\mathcal{O}(nm^n)$, which is 
prohibitive, especially for long strings and large alphabets. 
Therefore, there is a need to develop new, efficient mechanisms, which is done next. 
\end{rem}

\section{Differential privacy over a finite alphabet}
\label{sec:dp_nonmarkov_chain}
This section solves Problems~\ref{prob:offline_mechanism} and~\ref{prob:factored_mechanism}
in Subsections~\ref{sec:offline_non_markov} and~\ref{sec:online_non_markov}, respectively. 

\subsection{Offline Mechanism}\label{sec:offline_non_markov}

To develop an efficient offline privacy mechanism, we introduce an automaton, 
called the \emph{modified Hamming distance NFA}, which identifies all words that have a 
specific Hamming distance to an input word.

\begin{defn}[Modified Hamming Distance NFA]\label{def:Modified_Hamming_Distance_NFA}
    Fix an alphabet $\Sigma$ and a length $n\in\mathbb{N}^+$. For a word $x\in\Sigma^n$ 
    and a distance $j \in \mathbb{N}$, the modified Hamming Distance NFA (MNFA) is an NFA $A_{x,j}=(Q_{x,j},\Sigma,q^0,\delta,F_{n,j})$ such 
    that $\mathcal{L}(A_{x,j})$ is the set of all words of length~$n$ with Hamming distance from~$x$ equal to~$j$. Each state $q_i\in Q_{x,j}$ can transfer to another state by a \emph{policy} $\mu(\cdot,\cdot \mid q_i):Q_{x,j}\times\Sigma\rightarrow [0,1]$, which is a probabilistic mapping. Formally, we have    
        $\sum\limits_{q_{i+1},\sigma_{i+1}}\mu(q_{i+1},\sigma_{i+1} \mid q_i)=1$,
    where~$\mu(q_{i+1}, \sigma_{i+1} \mid q_i)$ is the probability that the input
    symbol~$\sigma_{i+1}$ causes a transition from~$q_i$ to~$q_{i+1}$. 
\end{defn}

\begin{figure}
\centering
\includegraphics[width=.40\textwidth]{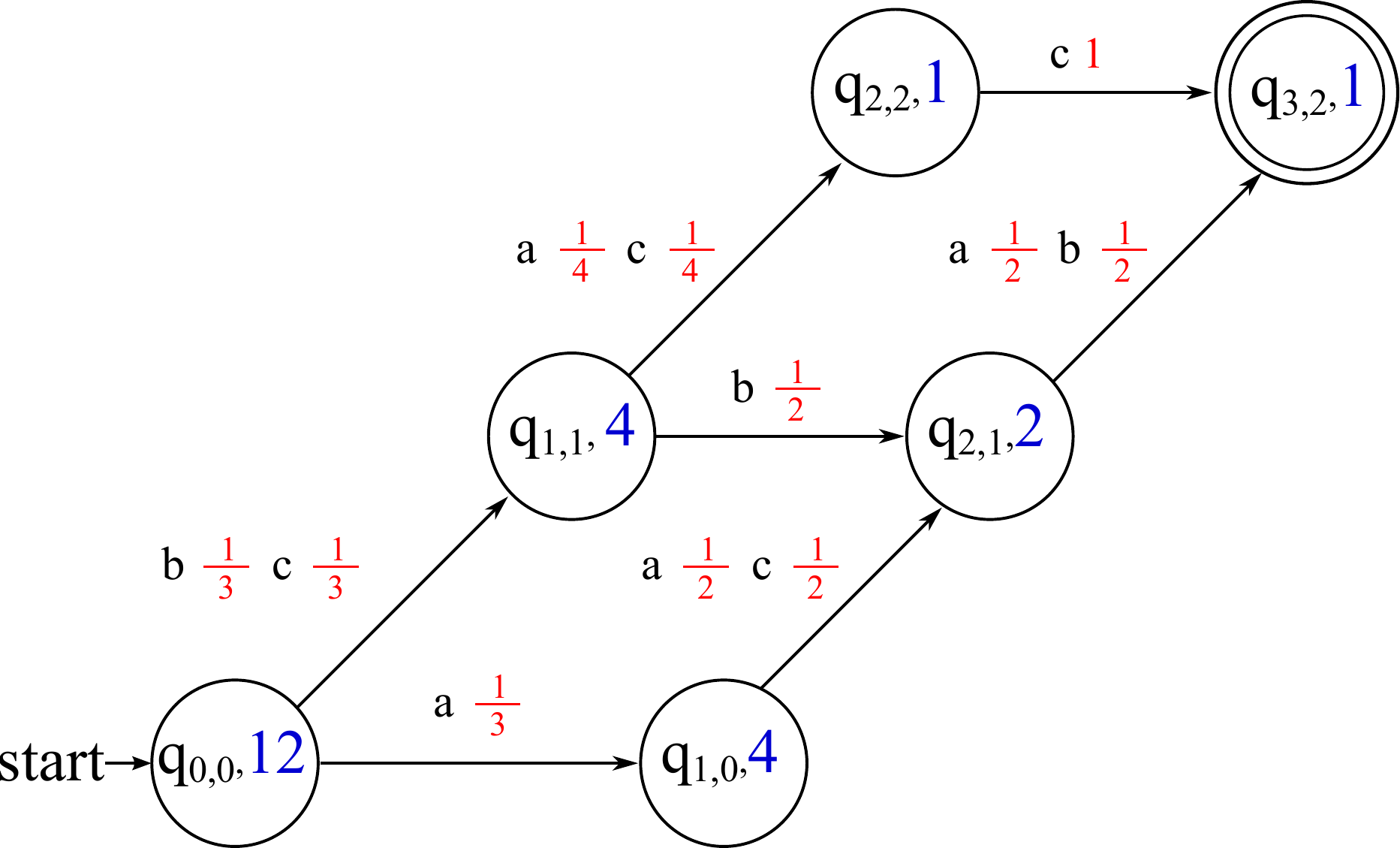}
\caption{Algorithm~\ref{alg:overall_offline_non_markov} for all words of length~$3$ and distance~$2$ from the word $abc$ over the alphabet
$\{a, b, c\}$. Each circle represents a state $q\in Q_{w_i,\ell}$, and each arrow represents a feasible transition. The state with 
a double circle denotes the accepting state $q_{n,\ell}$.
The value of $V(q)$ is in \textcolor{blue}{blue}. The value of $\mu_{\epsilon,w_i}$ is 
in \textcolor{red}{red} with the output letter in front of it.}
\label{fig:example_offline_mechanism}
\end{figure}

Definition~\ref{def:Modified_Hamming_Distance_NFA} is a modified form of the Hamming distance NFA
given in~\cite{Todd2017}.
Specifically, the definition in this paper 
has been modified so that all output words have Hamming distance~$j$ to the input word. 
Figure~\ref{fig:example_offline_mechanism} shows an example MNFA. The states are denoted in the form $q_{m,e}$, where~$m$
is the number of characters added to the output word so far, and~$e$ is the number of differences with the input
word that have been observed so far. 

Once the input word $x$ and a distance $j$ is given, we can construct the transition function by modifying the approach in~\cite{Schulz02faststring} for the Levenshtein automaton in a straight-forward way. 
Due to space constraints, we omit this construction here. 

We use the MNFA to develop an offline privacy mechanism as follows. 
Given an input word~$w_i$, the offline mechanism first randomly 
samples a Hamming distance~$\ell$, which will be the distance 
between~$w_i$ and its private 
output. Second, 
it constructs the transition function $\delta$. 
The offline mechanism then synthesizes a randomized policy over $\delta$, 
the procedure for which is 
shown in Algorithm~\ref{alg:overall_offline_non_markov}. Then the mechanism uses this 
policy to randomly select an output word to enforce word~$\epsilon$-differential privacy. 

\begin{mechanism}[Offline Mechanism]\label{mech:offline_non_markov_mechanism}
    Fix a probability space~$(\Omega, \mathcal{F}, \mathbb{P})$ and an
    adjacency parameter~$k \in \mathbb{N}$. 
    Given an alphabet~$\Sigma$ which contains~$m$ symbols 
    and a word~$w_i = \sigma_1 \cdots \sigma_n \in \Sigma^n$, define the offline 
    mechanism~$M_w^{\text{off}} : \Sigma^n \times \Omega \to \Sigma^n$, which chooses an output word~$w_o = \sigma_1^o \cdots \sigma_n^o \in \Sigma^n$ 
    by first drawing a distance $\ell$ from the distribution
    \begin{equation}\label{eq:probability_offline_mechanism}
        p(\ell;w_i,k)=\frac{\binom{n}{\ell}(m-1)^\ell\exp\left({-\frac{\epsilon\ell}{2k}}\right)}{\sum_{i=0}^{n}\binom{n}{i}(m-1)^i\exp\left({-\frac{\epsilon i}{2k}}\right)}.
    \end{equation}
    
    Then build a modified Hamming distance NFA $A_{w_i,k}=(Q_{w_i,k},\Sigma,q_{0,0},\delta,\{q_{n,\ell}\})$ by first constructing a transition function $\delta$ and then use Algorithm~\ref{alg:overall_offline_non_markov} to synthesize a policy. 
    An output word~$w_o=\sigma_1^o\dots\sigma_n^o\in\mathcal{L}(A_{w_i,k})$ is generated by running the NFA~$A_{w_i,k}$ once.
\end{mechanism}

\begin{algorithm2e}
\SetAlgoLined
\SetKwInOut{Input}{Input}
\SetKwInOut{Output}{Output}
\Input{Sensitive input string~$w_i$, transition function $\delta$, accepting set $\{q_{n,\ell}\}$}
\Output{Policy $\mu_{\epsilon,w_i}$}
 $n = |w_i|$\; 
$V(q_{n,\ell}) = 1$\;
$CurrQ = \{q_{n,\ell}\}$\;
$ActiveQ = \{\}$\;
$counter = 1$\;
\While{$counter\leq n$}{
    \For{$q'\in CurrQ$}{
        \For{$(q, \sigma)$ s.t. $q' \in \delta(q,\sigma)$}{
        $V(q) = \sum_{\{q'' \mid \exists \alpha, \,q'' \in \delta(q,\alpha)\}}V(q'')$\;
        $\mu_{\epsilon,w_i}(q',\sigma \mid q)=\frac{V(q')}{V(q)}$\;
        $ActiveQ = ActiveQ + \{q\}$\;}
    }
    $CurrQ = ActiveQ$\;
    $ActiveQ=\{\}$\;
    $counter=counter+1$\;
}
\caption{Modified Hamming Distance NFA Construction and Policy Synthesis}
 \label{alg:overall_offline_non_markov}
\end{algorithm2e}

Algorithm~\ref{alg:overall_offline_non_markov} 
takes as input a sensitive string~$w_i$, the transition function $\delta$, and the accepting set $\{q_{n,\ell}\}$. It outputs a 
policy $\mu_{\epsilon,w_i}$ which is designed as follows.
We assign a function $V:Q_{w_i,\ell}\to\mathbb{N}$ such that $V(q)$ is the number of unique 
paths from the state $q\in Q_{w_i,\ell}$ that end in $\{q_{n,\ell}\}$. The probability 
of outputting a symbol~$\sigma \in \Sigma$ 
at each state $q$ is equal to the proportion of $V(q)$ compared to $V(\delta(q,\sigma))$. 
This ensures that all potential private outputs words distance $\ell$ from $w_i$ are 
equiprobable.

Algorithm~\ref{alg:overall_offline_non_markov} is \emph{backward reachable}: 
it starts at an accepting state $q_{n,\ell}$ and sets~$V(q_{n,\ell})=1$ 
and~$CurrQ=\{q_{n,\ell}\}$. It then 
loops through all states that can reach any state in $CurrQ$, and 
finds the corresponding unique paths. 
These states are stored in the set~$ActiveQ$. At the end of each iteration, we 
set~$ActiveQ$ to be the new~$CurrQ$ set, and we reinitialize~$ActiveQ$. 
Then this process repeats until it reaches the initial state.

In Equation~\eqref{eq:probability_offline_mechanism},~$p(\ell;w_i,k)$ induces a probability 
distribution over output words, and Algorithm~\ref{alg:overall_offline_non_markov} provides 
the means to efficiently sample from it. An example output of Algorithm~\ref{alg:overall_offline_non_markov} 
is shown in Figure~\ref{fig:example_offline_mechanism}. 

\begin{thm}[Solution to Problem~\ref{prob:offline_mechanism}] \label{thm:offline_dp}
    Given an adjacency parameter~$k \in \mathbb{N}$, a privacy parameter~$\epsilon \geq 0$,     
    and
    a sensitive word~$w_i \in \Sigma^n$, 
    Mechanism~\ref{mech:offline_non_markov_mechanism} provides word $\epsilon$-differentially privacy to~$w_i$ with respect to the adjacency relation~$\textnormal{Adj}_{n,k}$     in Definition~\ref{def:word_adjacency}. 
\end{thm}
\emph{Proof:} See Appendix \ref{apdx:thm_1}. \hfill$\blacksquare$

For a Hamming distance $\ell$ generated with
Equation~\eqref{eq:probability_offline_mechanism}, the time complexity for an alphabet that contains $m$ symbols to generate a private output 
using Mechanism~\ref{mech:offline_non_markov_mechanism} is $\mathcal{O}(nm)$, which
is a significant improvement over the direct implementation of the exponential
mechanism (cf. Remark~\ref{rem:complexity}). 
This is possible because Algorithm~\ref{alg:overall_offline_non_markov} 
narrows down the set of possible private outputs significantly by restricting the potential 
outputs to those with distance exactly~$\ell$ from~$w_i$. 

There is also a need to analyze the accuracy of private words. 
Such analyses enable informed calibration of privacy by balancing the strength of privacy
protections with the errors they induce. 
This is done by computing the expectation and variance of the distance between
input and output words as a function of~$\epsilon$. 

\begin{thm}\label{thm:concentration_bounds_offline_non_markov}
    Fix an alphabet~$\Sigma$ that contains~$m$ symbols, a privacy parameter~$\epsilon \geq 0$, an adjacency
    parameter~$k \in \mathbb{N}$,
    and an input word~$w_i \in \Sigma^n$.
    Under Mechanism~\ref{mech:offline_non_markov_mechanism}, 
    the distance~$\ell$ to the private output satisfies 
    \begin{align}
        E[\ell] &= 
        n - \frac{n}{[(m-1)\exp(-\frac{\epsilon}{2k})+1]}\label{eq:expectation_offline_nonmarkov}\\        
        Var[\ell] &= \frac{n[(m-1)\exp(-\frac{\epsilon}{2k})]}{[(m-1)\exp(-\frac{\epsilon}{2k})+1]^2}.\label{eq:variance_offline_nonmarkov}
    \end{align}
    For any $\eta\in(0,0.5)$, we have the concentration bound
        $\mathbb{P}\big[\big|d(w_i,w_o)-E[d(w_i,w_o)]\big| > \eta\big] \leq 2\exp\left(-\frac{2\eta^2}{n^2}\right)$. 
\end{thm}
\emph{Proof:} See Appendix \ref{apdx:thm_2}. \hfill$\blacksquare$

\begin{rem}
    When $\epsilon\to 0$, we see that the expected value $E[\ell] \to n - \frac{n}{m}$ 
    and~$Var[\ell] \to \frac{n(m - 1)}{m^2}$.
    Then highly private outputs are random strings that differ
    from their inputs in nearly every entry.     
    On the other hand, as $\epsilon\to\infty$, 
    $E[\ell]=0$ and $Var[\ell]=0$. Thus, as privacy vanishes, the mechanism
    simply outputs the input word unchanged.      
\end{rem}

\subsection{Online Mechanism}\label{sec:online_non_markov}
Future states are unknown in the online setting, and we propose an alternative mechanism: 
each output character is generated only based on the most recently generated sensitive 
symbol in a word. The correct output character (correct transition, CT) is assigned a
probability~$\tau$ that is controlled by the privacy parameter~$\epsilon$, and other symbols (substitutions, ST) 
are set to be equiprobable. 

\begin{mechanism}[Online Mechanism]\label{mech:online_non_markov_mechanism}
    Fix a probability space $(\Omega, \mathcal{F}, \mathbb{P})$. Given an alphabet $\Sigma$ that contains~$m$ symbols 
    and a word $w_i=\sigma_1\dots\sigma_n\in\Sigma^n$, the online mechanism $M_w^{\text{on}}$ chooses an output word $w_o=\sigma_1^o\dots\sigma_n^o\in\Sigma^n$ by selecting each $\sigma_t^o$ from the distribution $\mathbb{P}[\sigma_t^o]=\mu_\epsilon(\sigma_t^o \mid \sigma_t)$, 
    where $\mu_\epsilon(\cdot \mid \sigma_t):\Sigma\to[0,1]$ is a \emph{policy} 
    generated by Algorithm~\ref{alg:overall_online_non_markov}.
\end{mechanism}

\begin{algorithm2e}
\SetAlgoLined
\SetKwInOut{Input}{Input}
\Input{Sensitive word~$w_i$, probability of correct transition~$\tau$}
\SetKwInOut{Output}{Output}
\Output{Policy $\mu_\epsilon$}
\For{$\sigma_t\in\Sigma$}{
    \For{$\sigma_t^o\in\Sigma$}{
        \uIf{$\sigma_t^o=\sigma_t$}{
            $\mu_\epsilon(\sigma_t^o \mid \sigma_t) = \tau$\tcp*{CT}
        }
        \Else{
            $\mu_\epsilon(\sigma_t^o \mid \sigma_t) = \frac{1-\tau}{m-1}$\tcp*{ST}
        }
    }
}
\caption{Policy construction for online setting (Implements Mechanism~\ref{mech:online_non_markov_mechanism}; Solves to )}
\label{alg:overall_online_non_markov}
\end{algorithm2e}


At each time~$t$, the online mechanism~$M_w^{\text{on}}$ calls the policy~$\mu_\epsilon$,
which takes~$\sigma_t$ as input and outputs a probability distribution over~$\Sigma$ in terms of~$\tau$.
The value of~$\tau$ must be chosen to enforce differential privacy. 


\begin{thm}[Solution to Problem~\ref{prob:factored_mechanism}] \label{thm:correct_transition_non_markov}
    Fix an alphabet $\Sigma$ that contains $m$ symbols, a sensitive word~$w_i \in \Sigma^n$, a privacy parameter~$\epsilon \geq 0$, 
    and an adjacency parameter~$k \in \mathbb{N}$. 
    Then the online mechanism $M_w^{\text{on}}$ is word $\epsilon$-differentially 
    private with respect to the adjacency relationship~$\textnormal{Adj}_{n,k}$
    in Definition~\ref{def:word_adjacency} if $\tau$ in Algorithm~\ref{alg:overall_online_non_markov} satisfies
        $\tau = \frac{1}{(m-1)\exp\left(-\frac{\epsilon}{k}\right)+1}$.
\end{thm}

\emph{Proof:} See Appendix \ref{apdx:thm_3}. \hfill $\blacksquare$

For a privacy level $\epsilon$, one only needs to construct the policy $\mu$ 
once with time complexity $\mathcal{O}(m^2)$. Then every time the online 
mechanism receives an input symbol, one can call the policy $\mu_\epsilon$ 
to generate a private output symbol. 
The accuracy of Mechanism~\ref{mech:online_non_markov_mechanism} is quantified next. 

\begin{thm}\label{thm:concentration_bounds_online_non_markov}
    Fix an alphabet $\Sigma$ that contains~$m$ symbols, a privacy parameter~$\epsilon \geq 0$,
    and an adjacency parameter~$k \in \mathbb{N}$. 
    Suppose Mechanism~\ref{mech:online_non_markov_mechanism} takes~$w_i \in \Sigma^n$
    as input and generates~$w_o \in \Sigma^n$ as output. Then    
    \begin{align}
        &E[d(w_i,w_o)] = n - \frac{n}{[(m-1)\exp(-\frac{\epsilon}{k})+1]}\label{eq:expectation_online_nonmarkov},\\      
        &Var[d(w_i,w_o)]=\frac{n(m-1)\exp\left(-\frac{\epsilon}{k}\right)}{\big[(m-1)\exp\left(-\frac{\epsilon}{k}\right)+1\big]^2}\label{eq:variance_online_nonmarkov}.
    \end{align}
    For~$\eta \in (0, 1)$, we have     
    \begin{align*}
        &\mathbb{P}\big[d(w_i,w_o)>(1+\eta)E[d(w_i,w_o)]\big]\leq e^{-\frac{\eta^2}{2+\eta}E[d(w_i,w_o)]},\\
        &\mathbb{P}\big[d(w_i,w_o)<(1-\eta)E[d(w_i,w_o)]\big]\leq e^{-\frac{\eta^2}{2}E[d(w_i,w_o)]}.
    \end{align*}
\end{thm}
\emph{Proof:} See Appendix~\ref{apdx:thm_4}. \hfill$\blacksquare$

\begin{rem}
    A larger value of~$\epsilon$ gives weaker privacy protections.
    In Theorem~\ref{thm:concentration_bounds_online_non_markov},
    both the expectation and variance go to zero
    as~$\epsilon$ grows. 
    Thus, the online mechanism captures the intuition that
    weaker privacy should give output words
    closer to the input word. 
\end{rem}

For a symbolic system whose states take values in~$\Sigma$,
Mechanism~\ref{mech:offline_non_markov_mechanism} provides
the means to privatize state sequences in batches,
while Mechanism~\ref{mech:online_non_markov_mechanism} provides
the means to privatize state sequences as they are generated.
Both mechanisms generate outputs that take values in
the entirety of the state space~$\Sigma$, though some systems
have restrictions on which state-to-state transitions are feasible.
A privacy implementation for such a system must account for these
feasibility requirements, and the next section does
this for a class of systems. 

\section{Generating differentially private runs for a Markov Chain}\label{sec:dp_markov_chain}
This section solves Problem~\ref{prob:markov_chain}. 
That is, the principles used in Section~\ref{sec:dp_nonmarkov_chain}
to generate differentially private words from an arbitary alphabet 
are applied to Markov chains. Additional developments
are necessary for Markov chains because 
Mechanisms~\ref{mech:offline_non_markov_mechanism} 
and~\ref{mech:online_non_markov_mechanism} must be modified to 
account for the feasibility of transitions between states. 

For a Markov chain with state space~$S$, 
it is in general only possible to transition from
a given state~$s$ to a subset of other states in~$S$. 
In this section, we define the symbols
\begin{equation}
    C(s) = \big\{s'\in S \mid \mathbb{P}[s' \mid s]>0\big\}  \textnormal{ and } 
    N(s) = |C(s)|, \label{eq:feasible_state_number}
\end{equation}
i.e.,~$C(s)$ is all states that can be reached from~$s$,
and~$N(s)$ is the number of such states. 

\subsection{Offline Mechanism}\label{sec:offline_markov}
The offline mechanism for Markov Chains works similarly to general symbolic systems,
except that, to address the feasiblity problem, policies are
synthesized using a Product modified Hamming distance NFA. 

\begin{defn}[Product Modified Hamming Distance NFA]
    Let a Markov chain~$(S, P, s_0)$ be given. 
    For a sequence of states~$x\in S^n$ and a distance $j \in \mathbb{N}$, let $A_{x,j}=(Q_{x,j},\Sigma,q^0,\delta,F_{|x|,j})$ be a MNFA.
    Then the Product Modified Hamming Distance NFA (P-MNFA) is an MNFA~$A_{x,j,S}=(Q_S,\Sigma,q_{S}^{0},\delta_{S},F_{S})$, where
    \begin{multline}
        Q_S = Q \times S, \quad \delta_{S} : Q\times S \times \Sigma\to 2^{Q_S}, \quad q_S^0 = (q_0, s_0) \\
        \textnormal{and } F_{S} = \big\{(q_f,s)\in Q_S \mid q_f\in F_{n,j}, s\in S\big\},  
    \end{multline}
    and for any $(q',s')\in\delta_{S}(q,s,\sigma)$, we have $\delta(q,\sigma)=q'$ and $\mathbb{P}[s' \mid s]>0$. 
    A state~$q_s\in Q_S$ can transition to another state by a policy $\mu_s(\cdot,\cdot \mid q_s,s):Q_{w_i,k}\times S\rightarrow [0,1]$. $\mathcal{L}(A_{x,j,S})$ is the set of all feasible words of length~$n$ with Hamming distance from~$x$ equal to~$j$.
\end{defn}

In words,~$A_{x,j,S}$ is the synchronous product 
of~$A_{x,j}$ and the Markov chain~$(S, P, s_0)$, and
every product transition function~$\delta_S$ has to  
satisify both the transition function~$\delta$ and be a feasible
transition in the Markov chain.

\begin{mechanism}[Offline Mechanism for Markov Chains]\label{mech:offline_markov_mechanism}
    Fix a probability space $(\Omega, \mathcal{F}, \mathbb{P})$ and
    a Markov chain $(S, P, s_0)$. 
    Let an adjacency parameter~$k \in \mathbb{N}$ be given.
    Given a sensitive word $w_i = s_1s_2 \cdots s_n\in S^n$, define 
    the offline mechanism $M_{w,s}^{\text{off}}$ (the subscript $s$ indicates ``state dependent"),     
    which generates 
    an output word $w_o=s_1^os_2^o\dots s_n^o\in S^n$ by first 
    drawing a Hamming distance from the distribution
    \begin{equation}\label{eq:prob_off_mech_markov}
        p(\ell;w_i,k)=\frac{m_{\ell}\exp\left({-\frac{\epsilon\ell}{2k}}\right)}{\sum_{i=0}^{n}m_i\exp\left({-\frac{\epsilon i}{2k}}\right)},
    \end{equation}    
    where~$m_\ell$ is the number of feasible words in~$S^{n}$ with initial state~$s_0$ that are 
    distance~$\ell$ to the input word. After sampling 
    a distance, it constructs a P-MNFA $A_{x,k,S}=(Q_{w_i,k},S,(q_{0,0},s_0),\delta_s,\{(q_{n,\ell},s) \mid s\in S\})$ 
    and synthesizes a policy $\mu_{\epsilon,s}$ using 
    Algorithm~\ref{alg:overall_offline_markov}. An output 
    word~$w_o=s_1^o\dots s_n^o\in\mathcal{L}(A_{w_i,k,S})$ is 
    generated by running the NFA~$A_{w_i,k,S}$ once. 
\end{mechanism}

\begin{figure}
\captionsetup[subfigure]{justification=justified}
\centering
\begin{subfigure}{0.47\textwidth}
\centering
\includegraphics[clip,scale=.35]{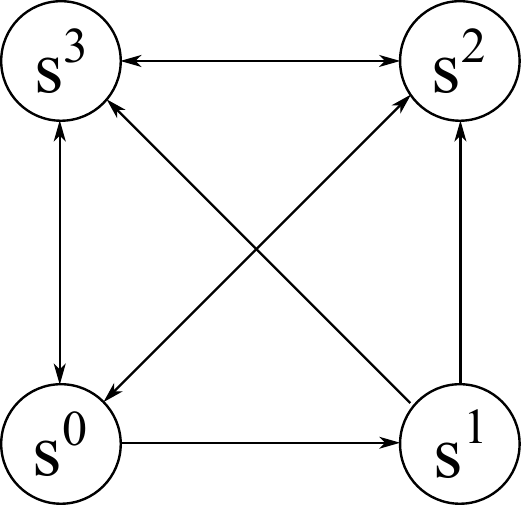}
\caption{A Markov chain with state space~$S=\{s^0,s^1,s^2,s^3\}$. 
Directed edges represent non-zero transition probabilities.}
\label{fig:Markov_chain_example}
\end{subfigure}

\begin{subfigure}{0.47\textwidth}
\centering
\includegraphics[clip,width=0.9\textwidth]{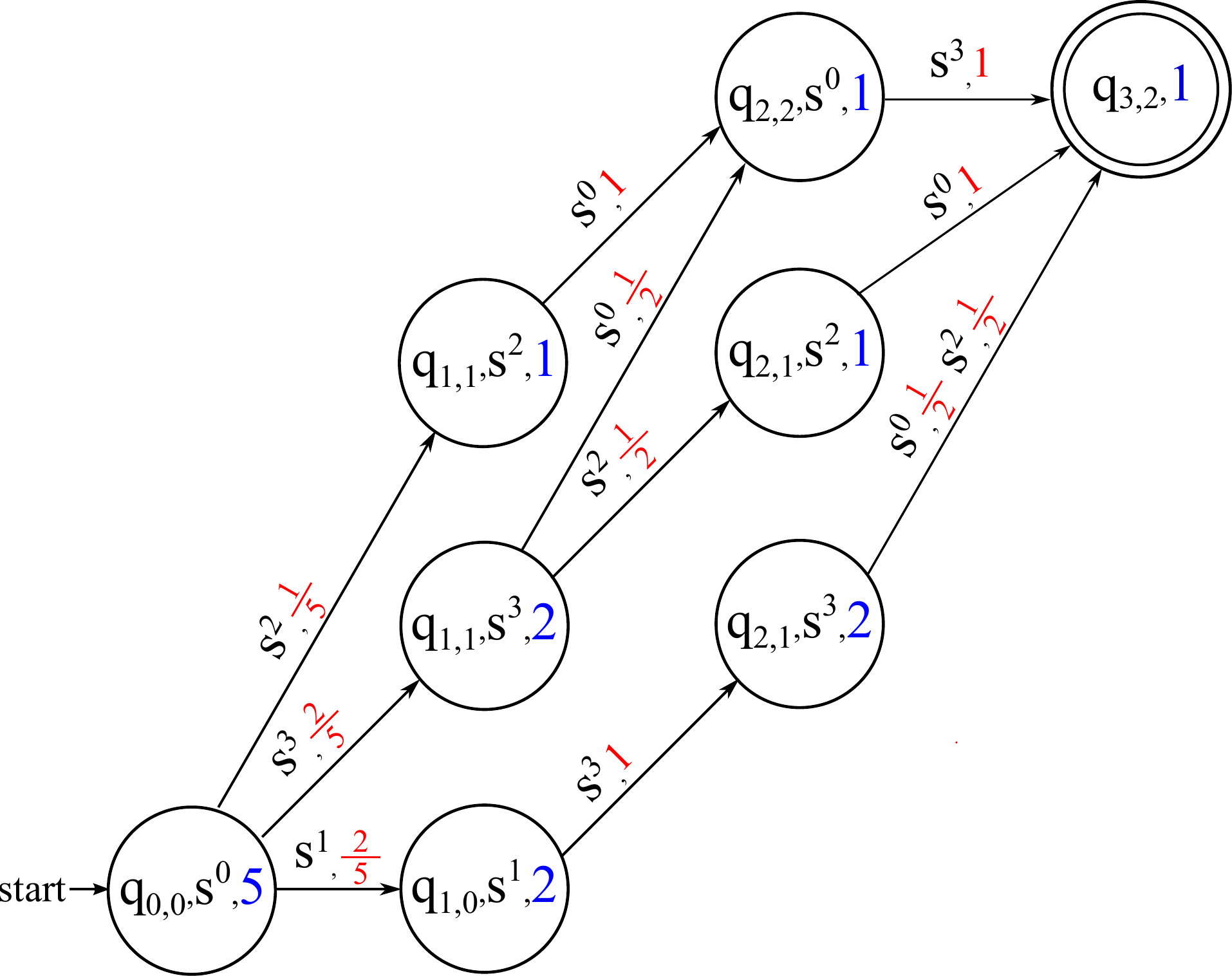}
\caption{Algorithm~\ref{alg:overall_offline_markov} for words of length 3 and distance 2 from the word~$s^1s^2s^3$ 
over the state space $\{s^0,s^1,s^2,s^3\}$. The output is started at $s^0$. The value of $V(q)$ is shown in \textcolor{blue}{blue}. 
The value of $\mu_{\epsilon,s}$ is shown in \textcolor{red}{red} with the input letter in front of it.}
\label{fig:example_offline_mechanism_Markov}
\end{subfigure}
\caption{Mechanism~\ref{mech:offline_markov_mechanism}
applied to the Markov chain in the upper figure leads to the MNFA in the lower figure.}
\label{fig:example_offline_markov}
\end{figure}

Mechanism~\ref{mech:offline_markov_mechanism} is similar to Mechanism~\ref{mech:offline_non_markov_mechanism}, but with the following differences. First, in Equation~\eqref{eq:prob_off_mech_markov}, $p(\ell;w_i,k)$ induces a probability distribution over only output words in $\mathcal{L}(S^{n})$. Second, observe that, in Mechanism~\ref{mech:offline_markov_mechanism},
for every output word~$w_o=s_1^os_2^o\dots s_n^o \in S^n$, 
feasibility always holds, i.e., $\mathbb{P}[s_{i+1}^o \mid s_i^o]>0$
for all~$i\in[n-1]$. This is because in Algorithm~\ref{alg:overall_offline_markov}, each state $(q,s)\in Q_S$ can transition 
to the state $\delta(q,s,s')$ only if $\mathbb{P}[s'\mid s]>0$. 
For a given Hamming distance $\ell$, the time complexity to generate a feasible private output is $\mathcal{O}(n|S|)$.
Figure~\ref{fig:example_offline_markov} shows an example use of 
Mechanism~\ref{mech:offline_markov_mechanism}.

\begin{algorithm2e}
\SetAlgoLined
\SetKwInOut{Input}{Input}
\SetKwInOut{Output}{Output}
\Input{Input string $w_i$, transition function $\delta_S$, accepting set $\{(q_{n,\ell},s) \mid s\in S\}$}
\Output{policy $\mu_{\epsilon,s}$}
$n = |w_i|$\; 
$V_s(q_{|w_i|,n},s) = 1,\forall s\in S$\;
$CurrQ = \{(q_{|w_i|,n},s)\},\forall s\in S$\;
$ActiveQ = \{\}$\;
$counter = 1$\;
\While{$counter\leq n$}{
    \For{$(q',s')\in CurrQ$}{
        \For{$(q,s)$ s.t. $\delta_{S}(q,s,s')=(q',s')$}{
        $V_s(q,s) = \sum_{\{(q'',s'') \mid \delta_{S}(q,s,s'')=(q'',s'')\}}V_s(q'',s'')$\;
        $\mu_{\epsilon,s}(q',s' \mid q,s)=\frac{V_s(q',s')}{V_s(q,s)}$\;
        $ActiveQ = ActiveQ + \{(q,s)\}$\;
    }
    }
    $CurrQ = ActiveQ$\;
    $ActiveQ=\{\}$\;
    $counter=counter+1$\;
}
 \caption{
 Product Modified Hamming Distance NFA Construction for Markov Chains
  }
 \label{alg:overall_offline_markov}
\end{algorithm2e}

\begin{thm}[Solution to first part of Problem~\ref{prob:markov_chain}]\label{thm:dp_offline_markov}
    Fix a privacy parameter~$\epsilon \geq 0$ and an adjacency parameter~$k \in \mathbb{N}$. 
    Suppose the state sequence~$w_i \in S^n$ is generated by the 
    Markov chain~$\mathcal{M} = (S, P, s_0)$. Then
    Mechanism~\ref{mech:offline_markov_mechanism} is word~$\epsilon$-differentially private
    with respect to the adjacency relation~$\textnormal{Adj}_{n,k}$,
    and the output word~$w_o$ is feasible for~$\mathcal{M}$. 
\end{thm}
\emph{Proof:} 
See Appendix \ref{apdx:thm_5}. 
\hfill$\blacksquare$

We next bound the error induced by Mechanism~\ref{mech:offline_markov_mechanism}. 

\begin{thm}\label{thm:concentration_bounds_offline_markov}
    Fix a state space~$S$, an adjacency parameter~$k \in \mathbb{N}$, and a privacy parameter~$\epsilon \geq 0$. 
    Define~$N_{max} = \max_{s \in S} N(s)$ and~$N_{min} = \min_{s \in S} N(s)$. 
    Then for an input word $w_i$, the distance 
    between $w_i$ and a private output word $w_o$ from Mechanism~\ref{mech:offline_markov_mechanism} obeys
    $\underline{E} \leq E[d(w_i,w_o)] \leq \overline{E}$ and
        $Var[d(w_i,w_o)] \leq \frac{n^2}{4}\label{eq:variance_bound_offline_markov}$,    
    where, using~$B_{\epsilon,k} = \exp(-\frac{\epsilon}{2k})$, we have 
    \begin{align}
        \underline{E}[d(w_i,w_o)] &=\\ &\frac{n(N_{min}-1)B_{\epsilon,k}[(N_{min}-1)B_{\epsilon,k} + 1]^{n-1}}{\sum_{i=0}^{n}m_i\exp\left({-\frac{\epsilon i}{2k}}\right)}\\
        \overline{E}[d(w_i,w_o)] &= \frac{nN_{max}B_{\epsilon,k}[N_{max}B_{\epsilon,k} \!+\! 1]^{n-1}}{\sum_{i=0}^{n}m_i\exp\left({-\frac{\epsilon i}{2k}}\right)}.
    \end{align}
\end{thm}
\emph{Proof:} See Appendix \ref{apdx:thm_6}.
\hfill$\blacksquare$

\subsection{Online Mechanism}\label{sec:online_markov}
For the online setting, Mechanism~\ref{mech:online_non_markov_mechanism} 
and Algorithm~\ref{alg:overall_online_non_markov} are modified
to only generate feasible words for a Markov chain. 

\begin{mechanism}[Online Mechanism for Markov Chains]\label{mech:online_markov_mechanism}
    Fix a probability space $(\Omega, \mathcal{F}, \mathbb{P})$ and a Markov chain $(S, P, s_0)$. 
    Given a word $w_i=s_0\dots s_n\in S^n$, define the online mechanism $M_{w,s}^{on}$ that chooses an 
    output word $w_o=s_0^o\dots s_n^o\in S^n$ such that each $s_t^o$ is selected from the 
    distribution $\mathbb{P}[s_t^o]=\mu_{\epsilon,s}(s_t^o \mid s_t,s_{t-1}^o)$, where $\mu_{\epsilon,s}$ 
    is the policy synthesized by Algorithm~\ref{alg:overall_online_markov}, and $N(s_{t-1}^o)$ is 
    defined in Equation~\eqref{eq:feasible_state_number}. In Algorithm~\ref{alg:overall_online_markov}, 
    \begin{align}
        &\beta(s_t,s_{t-1}^o) = \left\{\begin{array}{cc}
             1,& \text{if } \mathbb{P}[s_t \mid s_{t-1}^o]>0,  \\
             0,& \text{otherwise.}
        \end{array} \right.\label{eq:indicator}
    \end{align}
\end{mechanism}

\begin{algorithm2e}
\SetAlgoLined
\SetKwInOut{Input}{Input}
\SetKwInOut{Output}{Output}
\Input{Probability of correct transition $\tau$, indicator function $\beta$, number of feasible states $N(s)$ for each~$s \in S$}
\Output{$\mu_{\epsilon,s}$}
\For{$s_t\in S$}{
    \For{$s_{t-1}^o\in S$}{
        \For{$s_t^o\in S$}{
            \uIf{$s_t=s_{t}^o$ and  $\beta(s_t,s_{t-1}^o)=1$}{
                $\mu_{\epsilon,s}(s_t^o \mid s_t,s_{t-1}^o)= 
                \tau(s_t,s_{t-1}^o)$ \tcp*{CT}
            }
            \uElseIf{$s_t\neq s_{t}^o$ and $\beta(s_t^o,s_{t-1}^o)=1$}{
                $\mu_{\epsilon,s}(s_t^o \mid s_t,s_{t-1}^o) = \frac{1-\tau(s_t,s_{t-1}^o)\beta(s_t,s_{t-1}^o)}{N(s_{t-1}^o)-\beta(s_t,s_{t-1}^o)}$\tcp*{ST}
            }
            \Else{
                $\mu_{\epsilon,s}(s_t^o \mid s_t,s_{t-1}^o)=0$\;
            }
        }
    }
}
\caption{
Online Policy Construction for Product Modified Hamming Distance NFA
}
\label{alg:overall_online_markov}
\end{algorithm2e}

\begin{figure}
  \begin{subfigure}[b]{0.2\textwidth}
    \includegraphics[width=0.7\textwidth]{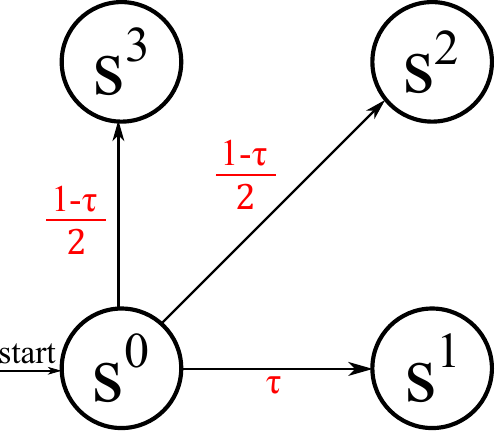}
    \caption{$\beta(s_t,s_{t-1}^o)=1$}
    \label{fig:online_markov_mechanism_ex_1}
  \end{subfigure}
  \hfill
  \begin{subfigure}[b]{0.2\textwidth}
    \includegraphics[width=0.7\textwidth]{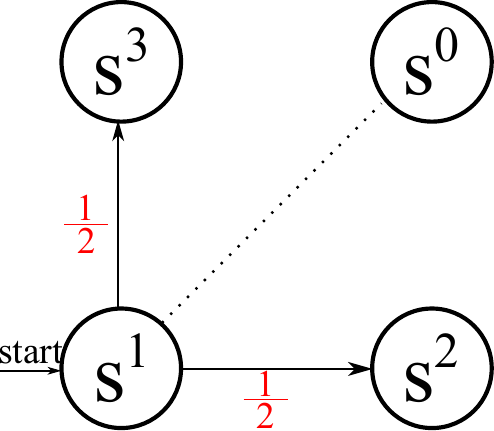}
    \caption{$\beta(s_t,s_{t-1}^o)=0$}
    \label{fig:online_markov_mechanism_ex_2}
  \end{subfigure}
  \caption{An example of Mechanism~\ref{mech:online_markov_mechanism}. The online mechanism $M_{\epsilon,s}^{on}$ is protecting a sensitive input state $s^1$ (i.e. $s_t=s^1$). Figure~\ref{fig:online_markov_mechanism_ex_1} shows the value of $\mu_{\epsilon,s}$ when the previous output state is $s^0$ (i.e. $s_{t-1}^o=s^0$) and Figure~\ref{fig:online_markov_mechanism_ex_2} shows when the previous state is $s^1$ (i.e. $s_{t-1}^o=s^1$). The value of $\mu_{\epsilon,s}$ is shown in \red{red} and the dashed edge indicates an infeasible transition. 
  }  
  \label{fig:online_markov_mechanism_ex}
\end{figure}

In words, at each time $t$, only the feasible states from the most recent output symbol~$s_{t-1}^o$
are assigned positive probabilities. If the current sensitive input character $s_t$ is feasible from $s_{t-1}^o$, 
then a correct transition is allowed with a probability $\tau(s_t,s_{t-1}^o)$ and other feasible states will have 
identical probabilities whose sum is $1-\tau(s_t,s_{t-1}^o)$. If $s_t$ is not feasible from $s_{t-1}^o$, the online 
mechanism cannot select the correct transition. Then its probability is set to $0$ and all feasible states 
are assigned equal probabilities. 
The time complexity of constructing the policy $\mu_{\epsilon,s}$ is $\mathcal{O}(|S|^3)$. 
After its construction, one can call this policy repeatedly to generate a private state. 
An example of Mechanism~\ref{mech:online_markov_mechanism} is shown in 
Figure~\ref{fig:online_markov_mechanism_ex}. 

\begin{thm}[Solution to second part of Problem~\ref{prob:markov_chain}] \label{thm:dp_online_Markov}
    Fix an adjacency parameter~$k\in \mathbb{N}$ and a privacy level~$\epsilon \geq 0$. 
    For a Markov chain~$(S, P, s_0)$, a sensitive input word $w_i = s_0\dots s_n\in S^n$ and an initial 
    state $s_0^o$, the online mechanism $M_{w,s}^{\text{on}}$ is word $\epsilon$-differentially 
    private with respect to the adjacency relationship~$\textnormal{Adj}_{n,k}$ 
    in Definition~\ref{def:word_adjacency} 
    if $\tau(s_t,s_{t-1}^o)$ in Algorithm~\ref{alg:overall_online_markov} satisfies
    \begin{equation}\label{eq:correct_transition_markov_online}
        \tau(s_t,s_{t-1}^o) = \frac{1}{(N(s_{t-1}^o)-1)\exp\left(-\frac{\epsilon}{k}\right)+1}.
    \end{equation}    
\end{thm}
\emph{Proof:} See Appendix \ref{apdx:thm_7}.\hfill$\blacksquare$

    For a given input word $w_i$, we can bound the expectation of the distance between $w_i$ and an private output word $w_o$ by substituting $N_{min}$ (or $N_{max}$) for $N(s_{t-1}^o)$ in Equation~\eqref{eq:correct_transition_markov_online}. The variance bound stays the same as in Theorem~\ref{thm:concentration_bounds_offline_markov}.

\begin{rem}
    The probability of a correct transition,~$\tau$, is a function of the most recent sensitive input symbol~$s_t$ 
    and most recent output $s_{t-1}^o$ instead of being a constant as in Algorithm~\ref{alg:overall_online_non_markov}. This is 
    because at different times the numbers of feasible states can
    be different. For example, in Figure~\ref{fig:Markov_chain_example}, if the initial 
    output state is $s_0$, then $C(s_0)=\{s_1,s_2,s_3\}$ and $N(s_0)=3$. But if the initial output state is $s_1$, 
    then $C(s_1)=\{s_1,s_2\}$ and $N(s_1)=2$. 
\end{rem}

\section{Simulation}\label{sec:experiment}
This section presents simulation results. Due to space constraints,
only examples of Mechanism~$4$ are shown. The example system we use is a Markov
chain model of a traffic system in Gainesville, FL that is generated by real-world traffic data. 
A symbolic trajectory produced by this Markov chain corresponds to a user's route
through Gainesville, and such routes are sensitive. For example, they may reveal
a user's place of home or work, their daily activities, and their acquaintances.
Therefore, we implement differential privacy for this system.

\begin{figure}[H]
\centering
\includegraphics[width=0.47\textwidth]{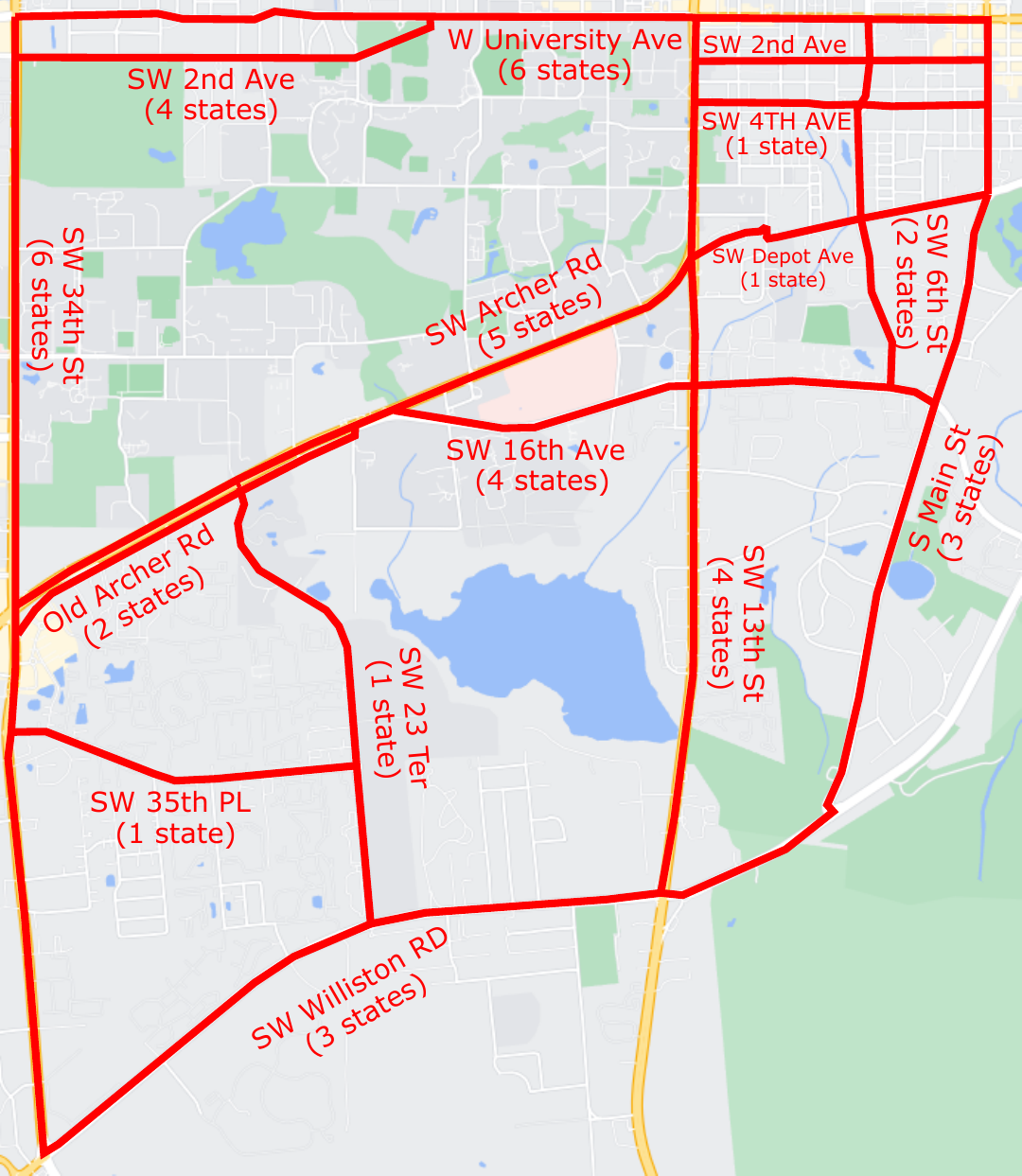}
\caption{Major streets in Gainesville, Florida, USA included in the Markov chain model used in this section. Each street is represented using one or more states,
and a user's symbolic state trajectory is the sequence of streets (or segments of streets)
that they travel along. 
}
\label{fig:Gainesville_street}
\end{figure}

\begin{figure}[H]
        \centering
        \includegraphics[width=0.47\textwidth]{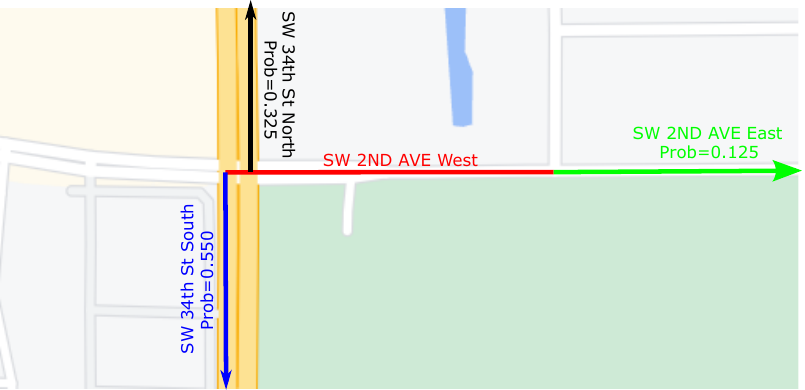}
        \caption{A portion of the Markov chain model used in this section. The red line (SW 2ND AVE West) is the initial state, and the arrows in different colors are 
        the states that it is feasible to transition to from the initial state, along with the probabilities of these transitions. 
        }
        \label{fig:Markov_street}
        \end{figure}

To elaborate, we consider a Markov chain that is generated by the Annual Average Daily Traffic (AADT) of some of the major streets in Gainesville, Florida from 2021. The traffic data can be obtained from Florida Traffic Online (2021)~\cite{florida2021}. Florida Traffic Online is a web-site mapping application which provides traffic count site locations and historical traffic count data. The AADT numbers are the total volumes of traffic on a road segment for one year, divided
by the number of days in that year. 
 The Markov chain we construct contains~$43$ states, where each state corresponds either to a single street or a segment of a street. We compute the transition probabilities in this model using basic frequency analysis. 
That is, the transition probability from one state to another is equal to the number
of times a driver transitions from the first state to the second divided by the total number of times
a driver transitions away from the first state. 
The alphabet $\Sigma$ contains one symbol for each state, and therefore $|\Sigma| = 43$. Figure~\ref{fig:Gainesville_street} shows all streets contained in the model, 
and Figure~\ref{fig:Markov_street} shows a portion of the Markov chain model along
with the relevant transition probabilities.

Next we demonstrate Mechanisms~\ref{mech:online_markov_mechanism} by generating differentially private versions of the route shown in Figure~\ref{fig:original}.

\subsection{Results for Different Privacy Parameters $\epsilon$}

\begin{figure*}[ht]
     \centering
     \begin{subfigure}[b]{0.3\textwidth}
        \centering
        \includegraphics[width=\textwidth]{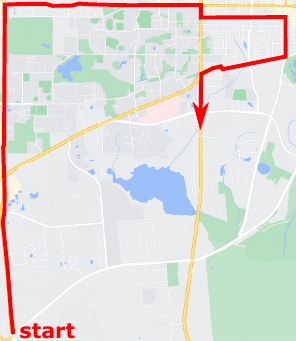}
        \caption{A car's sensitive route.}
        \label{fig:original}
    \end{subfigure}
     \hfill
     \begin{subfigure}[b]{0.3\textwidth}
        \centering
        \includegraphics[width=\textwidth]{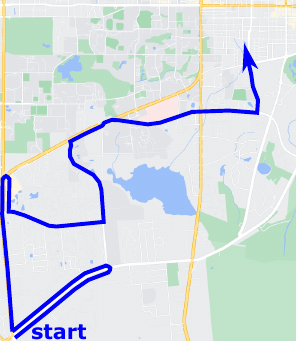}
        \caption{$\epsilon=0.1$, error$=14$}
    \end{subfigure}
    \hfill
     \begin{subfigure}[b]{0.3\textwidth}
        \centering
        \includegraphics[width=\textwidth]{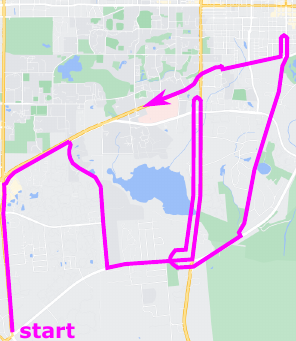}
        \caption{$\epsilon=1$, error$=11$}
    \end{subfigure}
    \\
    \begin{subfigure}[b]{0.3\textwidth}
        \centering
        \includegraphics[width=\textwidth]{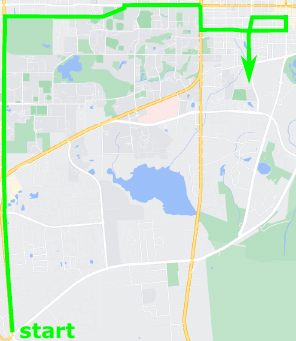}
        \caption{$\epsilon=3$, error$=5$}
    \end{subfigure}
     \hfill
     \begin{subfigure}[b]{0.3\textwidth}
        \centering
        \includegraphics[width=\textwidth]{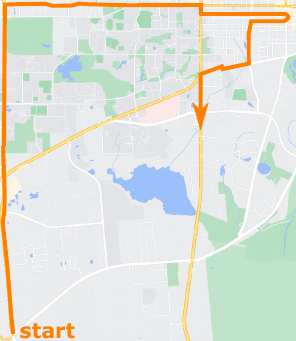}
        \caption{$\epsilon=5$, error$=1$}
    \end{subfigure}
    \hfill
     \begin{subfigure}[b]{0.3\textwidth}
        \centering
        \includegraphics[width=\textwidth]{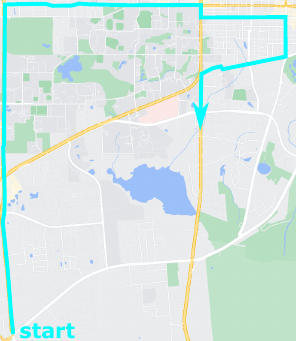}
        \caption{$\epsilon=10$, error$=0$}
    \end{subfigure}
    \caption{Differentially private samples of a car's route.
    Figure~\ref{fig:original} shows the sensitive route itself, and each other
    subfigure shows a private sample of that route, with the value of~$\epsilon$
    and number of errors shown in its corresponding caption. 
    All routes start from the initial state ``SW 34th St".}
    \label{fig:route}
\end{figure*}

We illustrate the effects of privacy parameters in the range $\epsilon\in[0.1, 10]$. 
Let the adjacency parameter $k = 1$. 
Figure~\ref{fig:route} gives example private outputs for different values of $\epsilon$. As $\epsilon$ grows, there is a general decrease in the distance between the sensitive input
route, shown in Figure~\ref{fig:original}, and the private outputs, shown in the
other subfigures. The private routes become more recognizable as~$\epsilon$ grows, and 
this agrees with intuition because a larger $\epsilon$ implies weaker privacy protections and thus should provide private outputs closer to the inputs that produced them.

\subsection{Results for Different Initial States}

\begin{figure}
\centering
\includegraphics[width=0.47\textwidth]{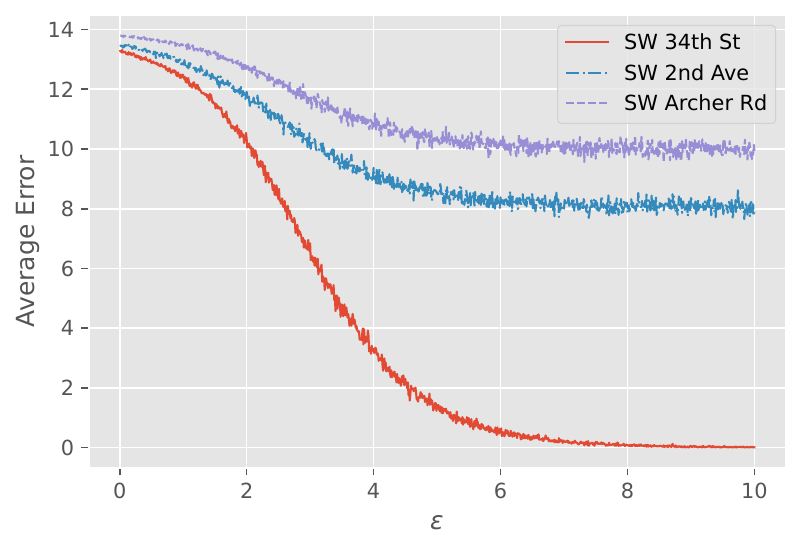}
\caption{Average distance between inputs and outputs from the initial states ``SW 34th St”, ``SW 2nd Ave”, and “SW Archer Rd”. With a larger $\epsilon$, which gives weaker privacy, the average error
decreases quickly. However, when $\epsilon\rightarrow\infty$, average errors of different initial states converge to different values because they deviate from the correct route in varying degrees.
The sensitive route length (excluding the fixed initial state) is 14.}
\label{fig:error_epsilon_Gainesville_street}
\end{figure}

We now consider different initial conditions of the Markov chain and explore
how they affect the accuracy of private trajectories. 
Let the adjacency parameter $k = 1$. 
 Figure~\ref{fig:error_epsilon_Gainesville_street} shows the average error between sensitive input words and 
private output words that start from three different initial states. 
For each initial state,~$1,000$ private output words were generated for each value of~$\epsilon$ 
to compute the average error. 
Figure~\ref{fig:error_epsilon_Gainesville_street} shows that for different initial conditions, Mechanism~\ref{mech:online_markov_mechanism} will generate outputs with different average errors. Specifically, at the same level of privacy, the online mechanism starting at the state “SW 34th St” tends to make fewer errors than when starting at the state “SW Archer Rd”. This is because the state “SW Archer Rd” deviates from the correct string.

To explore this further, let $s_t^o$ denote the random variable that is the output symbol at time~$t$, and let~$d_t$ denote the distance between the 
input symbol at time~$t$ and~$s^o_t$. 
Fix~$\epsilon = 5$. Then for different choices of~$s_0^o$ we list the probability of a correct transition at each time~$t \in [5]$ in Table~\ref{table:correct_transition}, i.e., we list the
probability that the output symbol~$s^o_t$ is equal to the input symbol at time~$t$.

\begin{table}
\centering
\begin{tabular}{ |c|m{5em}|m{7em}| } 
 \hline
 \diagbox[width=9em]{$t$}{Initial State} & SW 34th St & SW Archer Rd\\ 
 \hline
 1& 0.935& 0\\
 \hline
 2 & 0.993 & 0.245\\ 
 \hline
 3 & 0.967 & 0.242\\ 
 \hline
 4& 0.954 & 0.240\\
 \hline
 5& 0.948 & 0.237\\
 \hline
\end{tabular}
\caption{Probability of a correct transition at times $t\in[5]$ starting from the
initial states ``SW 34th St" and ``SW Archer Rd". Here~$\epsilon = 5$.}
\label{table:correct_transition}
\end{table}

For each time~$t$, the probability of a correct transition when starting at the state “SW 34th St” is high $(>0.9)$. As for the initial state “SW Archer Rd”, the Markov chain
dynamics force Mechanism~\ref{mech:online_markov_mechanism} to begin with an incorrect transition (i.e.,~$s^o_1$ must
be different from the input symbol at time~$1$) because the correct transition is infeasible.
We also see that beginning from “SW Archer Rd” gives Mechanism~\ref{mech:online_markov_mechanism} 
a lower chance to make correct transitions later. This is because the mechanism assigns every feasible state the same probability of transition when the mechanism cannot make the correct transition, even though transitioning to some states would lead to fewer errors in the long run. This assignment of probabilities is done because future states cannot be known at the present time, and the errors here are due to the lack of knowledge of the future that is inherent
to the online setting.

\section{Conclusion}\label{sec:conclusion}
This paper presented a 
novel differential privacy framework for symbolic systems that generate sensitive non-numerical sequences. 
Differential privacy is enforced by efficient mechanisms, and these can
be implemented both offline and 
online. 
These mechanisms were also extended to Markov Chains, and concentration bounds
were presented to quantify the accuracy of private outputs.
Future work will apply these mechanisms to Markov decision processes. By doing so, 
an MDP becomes a partially observable MDP (POMDP),
and future work will explore
privacy and performance in reinforcement learning problems modeled with MDPs.

\bibliographystyle{ieeetr}        
\bibliography{root}           



\begin{appendices}

\section{Proof of Theorem~\ref{thm:offline_dp}}\label{apdx:thm_1}
We proceed by showing that Mechanism~\ref{mech:offline_non_markov_mechanism} implements the exponential mechanism in Definition~\ref{def:exponential_mechanism} with the utility function in Definition~\ref{def:utility_function}. 
The exponential mechanism outputs the word~$w_o\in\Sigma^n$ with probability
\begin{align}
    p_e(w_o)&=\frac{\exp\left(\frac{-\epsilon d(w_i,w_o)}{2k}\right)}{\sum\limits_{w'\in \Sigma^n}\exp\left(\frac{-\epsilon d(w_i,w')}{2k}\right)} 
    \\
    &= \frac{\exp\left({-\frac{\epsilon d(w_i,w_o)}{2k}}\right)}{\sum_{i=0}^{n}\binom{n}{i}(m-1)^i\exp\left({-\frac{\epsilon i}{2k}}\right)},\label{eq:theorem_1_proof_exponential}
\end{align}
where the second equation holds since for each distance $i$, there are $\binom{n}{i}(m-1)^i$ possible outputs.
For Mechanism~\ref{mech:offline_non_markov_mechanism}, 
the probability of outputting the same~$w_o$ is
\begin{align}
    p_o(w_o) &= p(\ell;w_i,d(w_i,w_o))\cdot \frac{1}{\binom{n}{d(w_i,w_o)}(m-1)^{d(w_i,w_o)}}\nonumber\\
    &= \frac{\exp\left({-\frac{\epsilon d(w_i,w_o)}{2k}}\right)}{\sum_{i=0}^{n}\binom{n}{i}(m-1)^i\exp\left({-\frac{\epsilon i}{2k}}\right)}.\label{eq:theorem_1_proof_offline}
\end{align}
The first equation holds since each possible output word that has the same distance to the input word is equalprobable. As $p_e(w_o)=p_o(w_o)$ and the exponential mechanism is word $\epsilon$-differentially private, we conclude that the offline mechanism is word $\epsilon$-differentially private.
\hfill$\blacksquare$

\section{Proof of Theorem~\ref{thm:concentration_bounds_offline_non_markov}}\label{apdx:thm_2}
By expanding~$E[\ell] = \sum_{\ell=0}^{n}p(\ell;w_i,k)\ell$ we have
\begin{equation}
    E[\ell] =\frac{\sum_{\ell=0}^{n}\ell\binom{n}{\ell}(m-1)^\ell\exp\left({-\frac{\epsilon\ell}{2k}}\right)}{\sum_{i=0}^{n}\binom{n}{i}(m-1)^i\exp\left({-\frac{\epsilon i}{2k}}\right)}\label{eq:theorem_2_proof_1},
\end{equation}
which follows by plugging in $p(\ell;w_i,k)$ from 
Equation~\eqref{eq:probability_offline_mechanism}.
The numerator of Equation~\eqref{eq:theorem_2_proof_1} is equal to
\begin{align}
    \sum_{\ell=0}^{n}&\ell\binom{n}{\ell}(m-1)^\ell\exp\left({-\frac{\epsilon\ell}{2k}}\right)   
    =n(m-1) \\
    &\cdot\exp\left(-\frac{\epsilon}{2k}\right)\left[(m \!-\! 1)\exp\left(-\frac{\epsilon}{2k}\right) \!+\! 1\right]^{n-1},  \label{eq:theorem_2_proof_3}   
\end{align}
which follows by factoring out~$n(m-1)\exp\left(-\frac{\epsilon}{2k}\right)$ and
using the binomial theorem on the resulting sum. The binomial theorem
can also be used for the denominator of Equation~\eqref{eq:theorem_2_proof_1}. 
That result and
Equation~\eqref{eq:theorem_2_proof_3} 
give the expectation.

For variance, expanding~$E[\ell^2] = \sum_{\ell=0}^{n}p(\ell;w_i,k)\ell^2$ gives
\begin{equation}
    E[\ell^2] = \frac{\sum_{\ell=0}^{n}\ell^2\binom{n}{\ell}(m-1)^\ell\exp\left({-\frac{\epsilon\ell}{2k}}\right)}{\sum_{i=0}^{n}\binom{n}{i}(m-1)^i\exp\left({-\frac{\epsilon i}{2k}}\right)}.\label{eq:theorem_2_proof_5}
\end{equation}
The denominator of Equation~\eqref{eq:theorem_2_proof_5} can be simplified using the binomial theorem. 
The numerator can be simplified by factoring out~$n(m-1)\exp\left(-\frac{\epsilon}{2k}\right)$, which
then gives
\begin{multline}
    \sum_{\ell=0}^{n}\ell^2\binom{n}{\ell}(m-1)^\ell\exp\left({-\frac{\epsilon\ell}{2k}}\right)\\   
    = n(m-1)\exp\left(-\frac{\epsilon}{2k}\right)\cdot[b_1+b_2],    
\end{multline}
where
    $b_1 = \sum_{j=0}^{n-1}j\binom{n-1}{j}(m-1)^j\exp\left({-\frac{\epsilon j}{2k}}\right)$ and
    $b_2 = \sum_{j=0}^{n-1}\binom{n-1}{j}(m-1)^j\exp\left({-\frac{\epsilon j}{2k}}\right)$.
Then~$b_1$ can be simplified the same way as Equation~\eqref{eq:theorem_2_proof_3}, and $b_2$ can be simplified using the bionomial theorem. Then~$E[\ell^2]$ is
\begin{equation}
n(m-1)\exp\left(-\frac{\epsilon}{2k}\right)\left(n(m-1)\exp\left(-\frac{\epsilon}{2k}\right) + 1\right),
\end{equation}
and we use~$Var[\ell] = E[\ell^2] - E[\ell]^2$. Concentration bounds follow from Chernoff-Hoeffding bounds~\cite{Hoeffding1963}.
\hfill$\blacksquare$

\section{Proof of Theorem ~\ref{thm:correct_transition_non_markov}}  \label{apdx:thm_3}  
The given value
of~$\tau$ satisfies $\tau>\frac{1-\tau}{m-1}$, which follows from
    $\tau-\frac{1-\tau}{m-1}= \frac{1-\exp(-\epsilon/k)}{(m-1)\exp(-\epsilon/k)+1}>0$.
Then for all $(w_i,w_i')\in \textnormal{Adj}_{n,k}$ and any output word $w_o$, we have
    $p(w_o;w_i) =  \left(\frac{1-\tau}{m-1}\right)^{d(w_i,w_o)}\tau^{n-d(w_i,w_o)}$,
because each output character is chosen independently. 
A similar statement holds for~$p(w_o;w_i')$. 
Let~$\tilde{d}=d(w_i,w_o)-d(w_i',w_o)$. 
From~$(w_i,w_i')\in \textnormal{Adj}_{n,k}$,
it follows that~$-k \leq\tilde{d}\leq k$. Then
\begin{align*}
    &\frac{p(w_o;w_i)}{p(w_o;w_i')} \!=\! \frac{ \left(\!\frac{1-\tau}{m-1}\!\right)^{d(w_i,w_o)}\!\tau^{n-d(w_i,w_o)}}{\left(\!\frac{1-\tau}{m-1}\!\right)^{d(w_i',w_o)}\!\tau^{n-d(w_i',w_o)}} 
    \!=\! \left(\!\frac{1 \!-\! \tau}{m \!-\! 1}\!\right)^{\tilde{d}}\tau^{-\tilde{d}} \\
    &= \left(\frac{1-\tau}{(m-1)\tau}\right)^{\tilde{d}} 
    \leq \left(\frac{1-\tau}{(m-1)\tau}\right)^{-k} = \exp(\epsilon),
\end{align*}
where the first inequality holds because~$\frac{1-\tau}{(m-1)\tau}<1$, and~$\tilde{d}\geq -k$. The final 
equality holds by plugging in~$\tau$. Showing~$\frac{p(w_o;w_i)}{p(w_o;w_i')}\geq\exp(-\epsilon)$ uses the same technique.
\hfill $\blacksquare$


\section{Proof of Theorem~\ref{thm:concentration_bounds_online_non_markov}}\label{apdx:thm_4}
In Mechanism~\ref{mech:online_non_markov_mechanism}, $d(w_i,w_o)$ has the distribution
    $\mathbb{P}[d(w_i,w_o)=\ell] = \binom{n}{\ell}(m-1)^\ell\left(\frac{1-\tau}{m-1}\right)^\ell\tau^{n-\ell}
    = \binom{n}{\ell}(1-\tau)^\ell\tau^{n-\ell}$, 
which is Binomial. 
Then the result follows from
    $E[d(w_i,w_o)] \!=\! n(1-\tau)$, 
    $Var[d(w_i,w_o)] \!=\! n\tau(1-\tau)$, 
and expanding~$\tau$. 
Concentration bounds follow from Chernoff bounds~\cite{mcdiarmid_1989}.  
\hfill $\blacksquare$

\section{Proof of Theorem~\ref{thm:dp_offline_markov}}\label{apdx:thm_5}
Like the proof of Theorem~\ref{thm:offline_dp} in Appendix~\ref{apdx:thm_1}, 
this proof proceeds by showing that Mechanism~\ref{mech:offline_markov_mechanism} 
implements the exponential mechanism in Definition~\ref{def:exponential_mechanism}. 
For the input word~$w_i$, 
we will verify the equality of the probability with which
the exponential mechanism outputs~$w_o$, which is~$p_e(w_o)$,
and the probability with which that same word is output by
Mechanism~\ref{mech:offline_markov_mechanism}, which is~$p_o(w_o)$. 

Let $F(s_0)\subseteq S^n$ be the set of all feasible words with initial state $s_0$.
Suppose that~$d(w_i,w_o)=\ell$. Then 
\begin{equation}
    p_e(w_o) = \frac{\exp\left(-\frac{\epsilon \ell}{2k}\right)}{\sum\limits_{w'\in F(s_0)}\exp\left(\frac{\epsilon d(w_i,w')}{2k}\right)} 
    = \frac{\exp\left({-\frac{\epsilon \ell}{2k}}\right)}{\sum_{i=0}^{n}m_i\exp\left({-\frac{\epsilon i}{2k}}\right)},
\end{equation}
where~$m_i$ is the number of words in~$F(s_0)$ that are distance~$i$
from~$w_i$.

Mechanism~\ref{alg:overall_offline_markov} outputs the same~$w_o$
with probability 
\begin{equation}
    p_o(w_o) = p(\ell;w_i,k)\cdot \frac{1}{m_\ell} 
    = \frac{\exp\left({-\frac{\epsilon \ell}{2k}}\right)}{\sum_{i=0}^{n}m_i\exp\left({-\frac{\epsilon i}{2k}}\right)} =p_e(w_o).
\end{equation}
The first equation holds since each possible output word that has the same distance to the input word is equalprobable. 
The second equation expands~$p(\ell; w_i, k)$ and simplifies.
Since the exponential mechanism is word $\epsilon$-differentially private,  Mechanism~\ref{mech:offline_markov_mechanism} is, too.
\hfill $\blacksquare$

\section{Proof of Theorem~\ref{thm:concentration_bounds_offline_markov}}\label{apdx:thm_6}
Given~$w_i\in S^n$, we have
    $E[d(w_i,w_o)]=\frac{\sum_{\ell=0}^{n}\ell m_\ell\exp\left({-\frac{\epsilon \ell}{2k}}\right)}{\sum_{i=0}^{n}m_i\exp\left({-\frac{\epsilon i}{2k}}\right)}$.
For every~$\ell$,~$\binom{n}{\ell}(N_{min}-1)^\ell\leq m_\ell\leq \binom{n}{\ell}N_{max}^\ell$.
Then
\begin{align}
    &\frac{\sum_{\ell=0}^{n}\ell \binom{n}{\ell}(N_{min}-1)^\ell\exp\left({-\frac{\epsilon \ell}{2k}}\right)}{\sum_{i=0}^{n}m_i\exp\left({-\frac{\epsilon i}{2k}}\right)}\nonumber\\
    &\leq E[d(w_i,w_o)]\leq\frac{\sum_{\ell=0}^{n}\ell \binom{n}{\ell}N_{max}^\ell\exp\left({-\frac{\epsilon \ell}{2k}}\right)}{\sum_{i=0}^{n}m_i\exp\left({-\frac{\epsilon i}{2k}}\right)}.\label{eq:theorem_6_proof_1}
\end{align}
Equation~\eqref{eq:theorem_6_proof_1} can be further simplified using the same procedure 
to reach Equation~\eqref{eq:theorem_2_proof_3}. 
The variance bound follows from Popoviciu's inequality~\cite{bhatia2000better}. \hfill $\blacksquare$

\section{Proof of Theorem~\ref{thm:dp_online_Markov}}\label{apdx:thm_7}
Let~$w_1=s_{11}s_{12}\dots s_{1n}\in S^n$ be a sensitive input word and let~$w_2=s_{21}s_{22}\dots s_{2n}\in S^n$ be another such that~$d(w_1,w_2)=k$. 
Consider~$M_{w,s}^{on}(w_1)=M_{w,s}^{on}(w_2)=w_o$ and $w_o=s_1^os_2^o\dots s_n^o\in S^n$. Then 
\begin{align}
    \frac{\mathbb{P}[M_{w,s}^{on}(w_1)=w_o]}{\mathbb{P}[M_{w,s}^{on}(w_2)=w_o]}
    &=\frac{\prod_{i=1}^{n}\mu_{\epsilon,s}(s_{i}^o \mid s_{1i},s_{i-1}^o)}{\prod_{j=1}^{n}\mu_{\epsilon,s}(s_j^o \mid s_{2j},s_{j-1}^o)}\nonumber\\
    &= \prod_{i\in B}\frac{\mu_{\epsilon,s}(s_{i}^o \mid s_{1i},s_{i-1}^o)}{\mu_{\epsilon,s}(s_i^o \mid s_{2i},s_{i-1}^o)},\label{eq:thm_6_proof_1}
\end{align}
where $B=\{j\in[n] \mid s_{1j}\neq s_{2j}\}$ and $|B|\leq k$. 
Note that in Algorithm~\ref{alg:overall_online_markov}, we have
\begin{equation}
    \mu_{\epsilon,s}(s_t^o \mid s_{t},s_{t-1}^o)=
    \begin{cases}
        \frac{1-\tau(s_t,s_{t-1}^o)}{N(s_{t-1}^o)-1} & \text{if } \beta(s_t,s_{t-1}^o)=1  \\
        \frac{1}{N(s_{t-1}^o)} & \text{otherwise}
    \end{cases}.
\end{equation}
We next show that Equation~\eqref{eq:correct_transition_markov_online}
gives~$\tau(s_t,s_{t-1}^o)\geq \frac{1}{N(s_{t-1}^o)}\geq \frac{1-\tau(s_t,s_{t-1}^o)}{N(s_{t-1}^o)-1}$. 
The first inequality follows from
\begin{align*}
    \tau(s_t,s_{t-1}^o)&-\frac{1}{N(s_{t-1}^o)}\\
    \quad\quad&=\frac{1}{(N(s_{t-1}^o)-1)\exp\left(-\frac{\epsilon}{k}\right)+1}-\frac{1}{N(s_{t-1}^o)}\\
    \quad\quad&=\frac{(N(s_{t-1}^o)-1)(1-\exp\left(-\frac{\epsilon}{k}\right))}{N(s_{t-1}^o)(1+(N(s_{t-1}^o)-1)\exp\left(-\frac{\epsilon}{k}\right))} \\
    &\geq 0.
\end{align*}
The second equation holds by regrouping the two terms and the inequality holds since $\exp\left(-\frac{\epsilon}{k}\right)\leq 1$. We next prove that $\frac{1}{N(s_{t-1}^o)}\geq \frac{1-\tau(s_t,s_{t-1}^o)}{N(s_{t-1}^o)-1}$. Because we have
\begin{align*}
    &\frac{1}{N(s_{t-1}^o)}- \frac{1-\tau(s_t,s_{t-1}^o)}{N(s_{t-1}^o)-1}\\
    &\quad=\frac{N(s_{t-1}^o)-1-N(s_{t-1}^o)+N(s_{t-1}^o)\tau(s_t,s_{t-1}^o)}{N(s_{t-1}^o)(N(s_{t-1}^o)-1)}\\
    &= \! \frac{N(s_{t-1}^o)\tau(s_t,s_{t-1}^o)-1}{N(s_{t-1}^o)(N(s_{t-1}^o)-1)}
    \!=\! \frac{\frac{(N(s_{t-1}^o)-1)(1-\exp\left(-\frac{\epsilon}{k}\right))}{(N(s_{t-1}^o)-1)\exp\left(-\frac{\epsilon}{k}\right)+1}}{N(s_{t-1}^o)(N(s_{t-1}^o)-1)} \!\geq 0. 
\end{align*}
The second equation holds by factoring and the third holds by plugging in~$\tau(s_t,s_{t-1}^o)$. The final inequality holds since $\exp\left(-\frac{\epsilon}{k}\right)\leq 1$. Now we bound
Equation~\eqref{eq:thm_6_proof_1} with
\begin{multline}
    \prod_{i\in B}\frac{\mu_{\epsilon,s}(s_{i}^o \mid s_{1i},s_{i-1}^o)}{\mu_{\epsilon,s}(s_i^o \mid s_{2i},s_{i-1}^o)}
    \leq \prod_{i\in B}\frac{\tau(s_{1i},s_{i-1}^o)}{\frac{1-\tau(s_{2i},s_{i-1}^o)}{N(s_{2i},s_{i-1}^o)-1}} \\
    =\prod_{i\in B}\exp\left(\frac{\epsilon}{k}\right) 
    =\exp(\epsilon)\label{eq:thm_6_proof_3}.
\end{multline}
Equation~\eqref{eq:thm_6_proof_3} holds by expanding~$\tau$ at each 
state using~$|B|\leq k$. The bound
    $\prod_{i\in B}\frac{\mu_{\epsilon,s}(s_{i}^o \mid s_{1i},s_{i-1}^o)}{\mu_{\epsilon,s}(s_i^o \mid s_{2i},s_{i-1}^o)} \geq \exp(-\epsilon)$
follows using the same technique.
\hfill $\blacksquare$
\end{appendices}
\end{document}